\documentclass[12pt,prd,showpacs,tightenlines,nofootinbib]{revtex4}
\usepackage{bm}
\usepackage{graphics}
\usepackage{rotating}
\usepackage{epsfig}
\begin{document}
\title{
Semileptonic decays of $\Lambda_b$ baryons in
the relativistic quark model}  
\author{R. N. Faustov}
\author{V. O. Galkin}
\affiliation{Institute of Informatics in Education, FRC CSC RAS,
  Vavilov Street 40, 119333 Moscow, Russia}

\begin{abstract}
Semileptonic $\Lambda_b$ decays are investigated in the framework of
the relativistic quark model based on the quasipotential approach and
the quark-diquark picture of baryons. The
decay form factors are expressed through the overlap integrals of the
initial and final baryon wave functions. All calculations are done
without employing nonrelativistic and heavy quark expansions. The
momentum transfer dependence of the decay form factors is explicitly
determined in the whole accessible kinematical range without any
extrapolations or model assumptions. Both the heavy-to-heavy
$\Lambda_b\to\Lambda_c\ell\nu_\ell$ and heavy-to-light $\Lambda_b\to p\ell\nu_\ell$
decay branching fractions  are calculated. The
results agree within error bars with the experimental value of the
branching fraction of  the $\Lambda_b\to\Lambda_c^+l^-\bar\nu_l$
decay. From the recent LHCb data on the ratio of the  branching
fractions of the heavy-to-light and heavy-to-heavy semileptonic $\Lambda_b$ decays
the ratio of the Cabbibo-Kobayashi-Maskawa matrix elements
$|V_{ub}|/|V_{cb}|$ is obtained. It is consistent  with the corresponding ratio determined from the inclusive $B$ meson decays. 
\end{abstract}


\maketitle

\section{Introduction}

In recent  years significant experimental progress has been achieved in
studying properties of heavy baryons. The masses of all ground states of
charmed and bottom baryons have been measured except $\Omega_b^*$
\cite{pdg}. Many decay channels of these baryons were observed and new
more precise data are expected in the near future, since heavy baryons are
copiously produced at the LHC.  Weak decays of bottom baryons can serve as
an additional source for the determination of the
Cabbibo-Kobayashi-Maskawa (CKM) matrix elements $|V_{cb}|$ and
$|V_{ub}|$. Such determination is particularly important since there
exists some tension between the values of these matrix elements
extracted from exclusive and inclusive bottom meson weak decays
\cite{pdg,belle,kkhm}.  Very recently the LHCb Collaboration \cite{lhcb}
reported the first measurement of the ratio of the heavy-to-light
semileptonic $\Lambda_b\to p l\nu_l$ and heavy-to-heavy semileptonic
$\Lambda_b\to\Lambda_c l\nu_l$ decay rates in the constrained
kinematical regions, thus providing data for the determination of the
ratio of the CKM matrix elements $|V_{ub}/|V_{cb}|$ from baryon decays.

In order to calculate weak decay rates of bottom baryons it is necessary to
determine the form factors which parametrize  the matrix elements of
the weak current between initial and final baryon states. These form
factors depend on the momentum transfer from the initial baryon to the
final baryon. In the case of semileptonic bottom baryon decays both to
heavy and light final baryons the momentum transfer squared $q^2$ varies in a rather broad kinematical range. Therefore it is very important to explicitly determine the $q^2$ dependence of decay form factors in the whole kinematical range.   

In this paper we study semileptonic $\Lambda_b$ decays in the
framework of the relativistic quark model based on the quasipotential
approach and QCD. This model was successfully applied for
investigating various meson properties \cite{mass,slbdecay}. The heavy
and strange baryon spectroscopy was studied in the relativistic
quark-diquark picture in Refs.~\cite{hbar,barregge}  where masses and
wave functions of the ground and excited baryon states were
obtained. We also calculated the decay rates of heavy-to-heavy
semileptonic baryon transitions \cite{hbardecay} using the heavy quark
expansion. Both infinitely  heavy quark limit and first order $1/m_Q$
corrections were considered. It was shown that our model satisfies all
model independent relations following from the heavy quark  symmetry
\cite{iwb}. Leading and subleading baryon Isgur-Wise functions were
determined. It was found that $1/m_Q$ corrections give larger
contributions to heavy baryon decay rates than for heavy meson decay
rates.  Indeed, for heavy meson decays the heavy quark effective
theory (HQET) $\bar\Lambda$ parameter is determined by the light quark
energy while for heavy baryon decays this  parameter is proportional
to the light diquark energy which is almost 2 times larger. Here we
calculate the weak decay form factors without employing the heavy
quark expansion. This allows us to improve previous results and to
consider simultaneously heavy-to-heavy and heavy-to-light semileptonic
$\Lambda_b$ decays. It is important to point out that our model
provides the explicit $q^2$ dependence of the weak decay form factors in the whole accessible kinematical range without additional model assumptions and extrapolations. We consistently take into account all  relativistic effects including transformations of the baryon wave functions from the rest to the moving reference frame and contributions of the intermediate negative-energy states.   

The paper is organized as follows. In the next section we briefly describe the relativistic quark-diquark picture of heavy baryons. Calculation of the weak current matrix elements between baryon states in the quasipotential approach is discussed in Sec.~III. Using this method in Sec.~IV we determine the heavy-to-heavy and heavy-to-light weak decay form factors in the whole accessible kinematical range. Semileptonic $\Lambda_b$ decay rates and other observables are calculated in Sec.~V  and compared with available experimental data and previous calculations. The determination of the ratio of the CKM matrix elements $|V_{ub}|$ and $|V_{cb}|$ from the recent LHCb data \cite{lhcb} is discussed. We give our conclusions in Sec..~VI. Explicit expressions for the decay form factors as overlap integrals of baryon wave functions are listed in Appendix.

\section{Relativistic quark-diquark picture of baryons}

We study the semileptonic decays of $\Lambda_Q$ baryons in the relativistic quark-diquark picture in the
framework of the quasipotential approach.
 The interaction of two quarks in a diquark and the quark-diquark interaction  in a baryon are described by the
diquark wave function $\Psi_{d}$ of the bound quark-quark state
and by the baryon wave function $\Psi_{B}$ of the bound quark-diquark
state,  which satisfy the relativistic
quasipotential equation of the Schr\"odinger type \cite{mass}
\begin{equation}
\label{quas}
{\left(\frac{b^2(M)}{2\mu_{R}}-\frac{{\bf
p}^2}{2\mu_{R}}\right)\Psi_{d,B}({\bf p})} =\int\frac{d^3 q}{(2\pi)^3}
 V({\bf p,q};M)\Psi_{d,B}({\bf q}),
\end{equation}
where the relativistic reduced mass is
$$
\mu_{R}=\frac{M^4-(m^2_1-m^2_2)^2}{4M^3}, $$
and the center-of-mass system
relative momentum squared on mass shell is
$${b^2(M) }
=\frac{[M^2-(m_1+m_2)^2][M^2-(m_1-m_2)^2]}{4M^2}.
$$
Here $M$ is the bound state diquark or baryon mass,
$m_{1,2}$ are the masses of  quarks ($q_1$ and $q_2$) which form
the diquark or of the  diquark ($d$) and  quark ($q$) which form
the baryon ($B$), and ${\bf p}$  is their relative momentum.

The kernel 
$V({\bf p,q};M)$ in Eq.~(\ref{quas}) is the quasipotential operator of
the quark-quark or quark-diquark interaction which is constructed with
the help of the
off-mass-shell scattering amplitude, projected onto the positive-energy states. We assume that the effective
interaction is the sum of the usual one-gluon exchange term and the mixture
of long-range vector and scalar linear confining potentials, where
the vector confining potential contains the Pauli term. The resulting
quasipotentials  are given by the following expressions. \\
The quark-quark ($qq$) interaction in the diquark is
 \begin{eqnarray}
\label{qpot}
V({\bf p,q};M)&=&\bar{u}_{1}(p)\bar{u}_{2}(-p)\frac12\Bigl[\frac43\alpha_sD_{ \mu\nu}({\bf
k})\gamma_1^{\mu}\gamma_2^{\nu}\cr&&+ V^V_{\rm conf}({\bf k})
\Gamma_1^{\mu}({\bf k})\Gamma_{2;\mu}(-{\bf k})+
 V^S_{\rm conf}({\bf k})\Bigr]u_{1}(q)u_{2}(-q),
\end{eqnarray}\\
 The quark-diquark ($qd$) interaction in the baryon is
\begin{eqnarray}
\label{dpot}
\!\!\!\!\!\!\!\!V({\bf p,q};M)\!\!&=&\!\!\frac{\langle d(P)|J_{\mu}|d(Q)\rangle}
{2\sqrt{E_d(p)E_d(q)}} \bar{u}_{q}(p)  
\frac43\alpha_sD_{ \mu\nu}({\bf 
k})\gamma^{\nu}u_{q}(q)\cr
&&\!\!\!\!\!\!+\psi^*_d(P)\bar u_q(p)J_{d;\mu}\Gamma_q^\mu({\bf k})
V_{\rm conf}^V({\bf k})u_{q}(q)\psi_d(Q)+\psi^*_d(P)
\bar{u}_{q}(p)V^S_{\rm conf}({\bf k})u_{q}(q)\psi_d(Q), 
\end{eqnarray}
where $\alpha_s$ is the QCD coupling constant, $\langle
d(P)|J_{\mu}|d(Q)\rangle$ is the vertex of the 
diquark-gluon interaction which takes into account the diquark
internal structure and $J_{d;\mu}$ is the effective long-range vector
vertex of the diquark. The diquark momenta are $P=(E_d(p),-{\bf p})$, $Q=(E_d(q),-{\bf q})$ with $E_d(p)=\sqrt{{\bf p}^2+M_d^2}$. $D_{\mu\nu}$ is the  
gluon propagator in the Coulomb gauge, ${\bf k=p-q}$; $\gamma_{\mu}$ and $u(p)$ are 
the Dirac matrices and spinors, while $\psi_d(P)$ is the diquark wave
function. The factor 1/2 in the quark-quark  interaction
accounts for the difference of the colour factor compared to the
quark-antiquark case.

The effective long-range vector vertex of the quark is
defined by 
\begin{equation}
\Gamma_{\mu}({\bf k})=\gamma_{\mu}+
\frac{i\kappa}{2m}\sigma_{\mu\nu}\tilde k^{\nu}, \qquad \tilde
k=(0,{\bf k}),
\end{equation}
where $\kappa$ is the anomalous chromomagnetic moment of quarks.

In the nonrelativistic limit the vector and scalar confining
potentials  reduce to
\begin{eqnarray}
V^V_{\rm conf}(r)&=&(1-\varepsilon)(Ar+B),\qquad\nonumber\\
V^S_{\rm conf}(r)& =&\varepsilon (Ar+B),
\end{eqnarray}
where $\varepsilon$ is the mixing coefficient, and the
usual Cornell-like potential is reproduced
\begin{equation}
V(r)=-\frac43\frac{\alpha_s}{r}+Ar+B.
\end{equation}
Here we use the QCD coupling constant with freezing 
\begin{equation}
  \label{eq:alpha}
  \alpha_s(\mu^2)=\frac{4\pi}{\displaystyle\beta_0
\ln\frac{\mu^2+M_B^2}{\Lambda^2}}, \qquad \beta_0=11-\frac23n_f,
\qquad \mu=\frac{2m_1m_2}{m_1+m_2},
\end{equation} 
with the background mass $M_B=2.24\sqrt{A}=0.95$~GeV  and
$\Lambda=413$~MeV \cite{ltetr}.

All parameters of the model such as quark masses, parameters of the
linear confining potential $A$ and $B$, the mixing coefficient
$\varepsilon$ and anomalous chromomagnetic quark moment $\kappa$  were
fixed previously from calculations of meson and baryon properties
\cite{mass,hbar}. The constituent quark masses $m_u=m_d=0.33$
GeV, $m_c=1.55$ GeV, $m_b=4.88$ GeV and the parameters of the linear potential
$A=0.18$ GeV$^2$ and $B=-0.3$ GeV have the usual values of quark
models.  The value of the mixing coefficient of vector and scalar
confining potentials $\varepsilon=-1$ has been determined from the
consideration of the heavy quark expansion for the semileptonic heavy
meson decays and charmonium radiative decays \cite{mass}. The
universal Pauli interaction constant $\kappa=-1$  has been fixed from
the analysis of the fine splitting of heavy quarkonia ${}^3P_J$-
states \cite{mass}.  Note that the long-range chromomagnetic
contribution to the potential, which is proportional to $(1+\kappa)$,
vanishes for the chosen value of $\kappa=-1$.

\section{Matrix element of the weak current between baryon states}

To
calculate the heavy $\Lambda_Q$ baryon decay rate to the heavy or
light $\Lambda_q$ ($q=c$ or $u$, $\Lambda_u\equiv p$) baryon it is necessary to determine the
corresponding matrix element of the  weak current between baryon
states, which in the quasipotential approach is given by   
\begin{eqnarray}\label{mxet} 
&&\!\!\langle \Lambda_{q}(p_{{q}}) \vert J^W_\mu \vert \Lambda_Q(p_{Q})\rangle
=\int \frac{d^3p\, d^3q}{(2\pi )^6} \bar \Psi_{\Lambda_{q}\,{\bf p}_{{q}}}({\bf
p})\Gamma _\mu ({\bf p},{\bf q})\Psi_{\Lambda_Q\,{\bf p}_{Q}}({\bf q}),
\end{eqnarray}
where $\Gamma _\mu ({\bf p},{\bf
q})$ is the two-particle vertex function and  
$\Psi_{\Lambda\,{\bf p}}$ is the baryon  wave function projected onto the positive-energy states of 
quarks and boosted to the moving reference frame with momentum ${\bf
  p}$. The baryon wave function is the product of the diquark and
quark wave functions.

\begin{figure}
  \centering
  \includegraphics{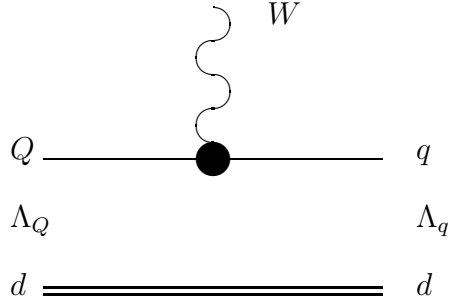}
\caption{Lowest order vertex function $\Gamma^{(1)}$
contributing to the current matrix element (\ref{mxet}). \label{d1}}
\end{figure}

\begin{figure}
  \centering
  \includegraphics{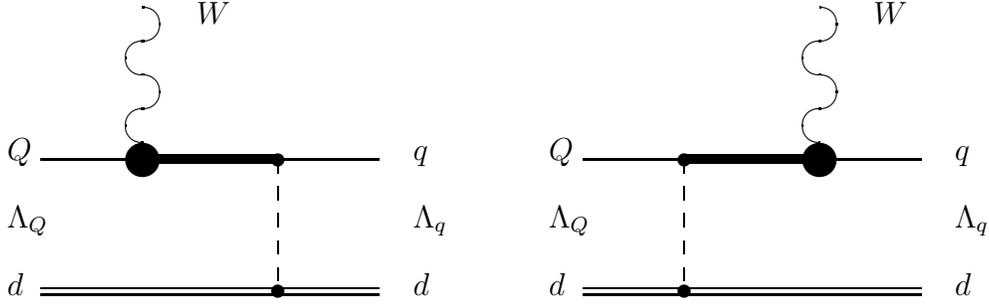}
\caption{ Vertex function $\Gamma^{(2)}$
taking the quark interaction into account. Dashed lines correspond  
to the effective potential ${\cal V}_{Qd}$ in 
(\ref{dpot}). Bold lines denote the negative-energy part of the quark
propagator. \label{d2}}
\end{figure}

 The contributions to $\Gamma$ come from Figs.~\ref{d1} and \ref{d2}. 
The contribution $\Gamma^{(2)}$ is the consequence
of the projection onto the positive-energy states. Note that the form
of the 
relativistic corrections resulting from the vertex function
$\Gamma^{(2)}$ is explicitly dependent on the Lorentz structure of the
quark-diquark interaction. For the heavy-to-heavy baryon transitions 
only $\Gamma^{(1)}$ contributes in the heavy quark limit ($m_{Q}\to \infty$), while $\Gamma^{(2)}$  
gives the subleading order contributions. 
The vertex functions are given by
\begin{equation} \label{gamma1}
\Gamma_\mu^{(1)}({\bf
p},{\bf q})=\psi^*_d(p_d)\bar
u_{q}(p_{q})\gamma_\mu(1-\gamma^5)u_Q(q_Q)\psi_d(q_d) 
(2\pi)^3\delta({\bf p}_d-{\bf
q}_d),\end{equation}
and
\begin{eqnarray}\label{gamma2} 
\Gamma_\mu^{(2)}({\bf
p},{\bf q})&=& \psi^*_d(p_d)\bar u_{q}(p_{q}) \Bigl\{\gamma_{\mu}(1-\gamma^5)
\frac{\Lambda_Q^{(-)}(
k)}{\epsilon_Q(k)+\epsilon_Q(p_{q})}\gamma^0
{\cal V}_{Qd}({\bf p}_d-{\bf
q}_d)\nonumber \\ 
& &+{\cal V}_{Qd}({\bf p}_d-{\bf
q}_d)\frac{\Lambda_{q}^{(-)}(k')}{ \epsilon_{q}(k')+
\epsilon_{q}(q_Q)}\gamma^0 \gamma_{\mu}(1-\gamma^5)\Bigr\}u_Q(q_Q)
\psi_d(q_d),\end{eqnarray}
where the superscripts ``(1)" and ``(2)" correspond to Figs.~\ref{d1} and
\ref{d2},  ${\bf k}={\bf p}_{q}-{\bf\Delta};\
{\bf k}'={\bf q}_Q+{\bf\Delta};\ {\bf\Delta}=
{\bf P}_{q}-{\bf P}_{Q}$; \ $\epsilon(p)=\sqrt{m^2+{\bf p}^2}$; and
$$\Lambda^{(-)}(p)=\frac{\epsilon(p)-\bigl( m\gamma
^0+\gamma^0({\bm{ \gamma}{\bf p}})\bigr)}{ 2\epsilon (p)}.$$

The wave functions in the weak current
matrix element (\ref{mxet}) are not in the rest frame. In the $\Lambda_Q$ baryon rest frame, the final  baryon
is moving with the recoil momentum ${\bf \Delta}$. The wave function
of the moving  baryon $\Psi_{{\Lambda_{q}}\,{\bf\Delta}}$ is connected 
with the  wave function in the rest frame 
$\Psi_{{\Lambda_{q}}\,{\bf 0}}\equiv \Psi_{\Lambda_{q}}$ by the transformation \cite{f}
\begin{equation}
\label{wig}
\Psi_{{\Lambda_{q}}\,{\bf\Delta}}({\bf
p})=D_{q}^{1/2}(R_{L_{\bf\Delta}}^W)D_d^{\cal I}(R_{L_{
\bf\Delta}}^W)\Psi_{{B_{q}}\,{\bf 0}}({\bf p}),\qquad  {\cal I}=0,1,
\end{equation}
where $R^W$ is the Wigner rotation, $L_{\bf\Delta}$ is the Lorentz boost
from the baryon rest frame to a moving one, and   
the rotation matrix of the quark spin $D^{1/2}(R)$ in the spinor
representation is given by 
\begin{equation}\label{d12}
{1 \ \ \,0\choose 0 \ \ \,1}D^{1/2}_{q}(R^W_{L_{\bf\Delta}})=
S^{-1}({\bf p}_{q})S({\bf\Delta})S({\bf p}),
\end{equation}
where
$$
S({\bf p})=\sqrt{\frac{\epsilon(p)+m}{2m}}\left(1+\frac{\bm{\alpha}{\bf p}}
{\epsilon(p)+m}\right)
$$
is the usual Lorentz transformation matrix of the four-spinor. The
rotation matrix $D^{\cal I}(R)$ of the diquark with spin ${\cal I}$ is
equal to $D_d^0(R^W)=1$ for the
scalar diquark and $D_d^1(R^W)=R^W$ for 
the axial vector diquark.

\section{Form factors  of the $\Lambda_b$ baryon decays}

The hadronic matrix elements of the vector and axial vector weak currents for the semileptonic decay $\Lambda_Q\to
\Lambda_{q}$  ($Q=b$ and $q=c$ or $u$) are parametrized  in terms of six invariant form factors:
\begin{eqnarray}
  \label{eq:llff}
  \langle \Lambda_{q}(p',s')|V^\mu|\Lambda_Q(p,s)\rangle&=& \bar
  u_{\Lambda_{q}}(p',s')\Bigl[F_1(q^2)\gamma^\mu+F_2(q^2)\frac{p^\mu}{M_{\Lambda_Q}}+F_3(q^2)\frac{p'^\mu}{M_{\Lambda_q}}\Bigl]
u_{\Lambda_Q}(p,s),\cr
 \langle \Lambda_{q}(p',s')|A^\mu|\Lambda_Q(p,s)\rangle&=& \bar
  u_{\Lambda_{q}}(p',s')\Bigl[G_1(q^2)\gamma^\mu+G_2(q^2)\frac{p^\mu}{M_{\Lambda_Q}}+G_3(q^2)\frac{p'^\mu}{M_{\Lambda_q}}\Bigl]
\gamma_5 u_{\Lambda_Q}(p,s),\qquad 
\end{eqnarray}
where   $u_{\Lambda_{Q}}(p,s)$ and
$u_{\Lambda_{q}}(p',s')$ are Dirac spinors of the initial and final
baryon; $q=p'-p$.

Another popular parametrization of these decay matrix elements reads \cite{giklsh,gikls}\begin{eqnarray}
  \label{eq:ff}
  \langle \Lambda_{q}(p',s')|V^\mu|\Lambda_Q(p,s)\rangle&=& \bar
  u_{\Lambda_{q}}(p',s')\Bigl[f_1^V(q^2)\gamma^\mu-f_2^V(q^2)i\sigma^{\mu\nu}\frac{q_\nu}{M_{\Lambda_Q}}+f_3^V(q^2)\frac{q^\mu}{M_{\Lambda_Q}}\Bigl]
u_{\Lambda_Q}(p,s),\cr
 \langle \Lambda_{q}(p',s')|A^\mu|\Lambda_Q(p,s)\rangle&=& \bar
  u_{\Lambda_{q}}(p',s')[f_1^A(q^2)\gamma^\mu-f_2^A(q^2)i\sigma^{\mu\nu}\frac{q_\nu}{M_{\Lambda_Q}}+f_3^A(q^2)\frac{q^\mu}{M_{\Lambda_Q}}\Bigl]
\gamma_5 u_{\Lambda_Q}(p,s),\qquad 
\end{eqnarray}
It is easy to find the following relations between these two sets of form factors:
\begin{eqnarray}
  \label{eq:rel}
  f_1^V(q^2)&=&F_1(q^2)+(M_{\Lambda_Q}+M_{\Lambda_q})\left[\frac{F_2(q^2)}{2M_{\Lambda_Q}}+\frac{F_3(q^2)}{2M_{\Lambda_q}}\right],\cr
f_2^V(q^2)&=&-\frac12\left[F_2(q^2)+\frac{M_{\Lambda_Q}}{M_{\Lambda_q}}F_3(q^2)\right],\cr
f_3^V(q^2)&=&\frac12\left[F_2(q^2)-\frac{M_{\Lambda_Q}}{M_{\Lambda_q}}F_3(q^2)\right],\cr
f_1^A(q^2)&=&G_1(q^2)-(M_{\Lambda_Q}-M_{\Lambda_q})\left[\frac{G_2(q^2)}{2M_{\Lambda_Q}}+\frac{G_3(q^2)}{2M_{\Lambda_q}}\right],\cr
f_2^A(q^2)&=&-\frac12\left[G_2(q^2)+\frac{M_{\Lambda_Q}}{M_{\Lambda_q}}G_3(q^2)\right],\cr
f_3^A(q^2)&=&\frac12\left[G_2(q^2)-\frac{M_{\Lambda_Q}}{M_{\Lambda_q}}G_3(q^2)\right].
\end{eqnarray}

To find the weak decay form factors we need to calculate the  matrix element of the weak current between baryon wave functions known from the mass spectra calculations. 
The general structure of the current matrix element (\ref{mxet}) is
rather complicated, because it is necessary to integrate both with
respect to $d^3p$ and $d^3q$. The $\delta$-function in the expression
(\ref{gamma1}) for the vertex function $\Gamma^{(1)}$ permits us to perform
one of these integrations. As a result the contribution of
$\Gamma^{(1)}$ to the current matrix element has the usual structure of
an overlap integral of baryon wave functions and
can be calculated exactly  in the
whole kinematical range. The situation with the contribution
$\Gamma^{(2)}$ is different. Here, instead of a $\delta$-function, we have
a complicated structure, containing the potential of the quark-diquark
interaction in the baryon. Therefore in the general case we cannot get rid of one
of the integrations in the contribution of $\Gamma^{(2)}$ to the
matrix element (\ref{mxet}). Thus it is necessary to use some 
additional considerations in order to simplify calculations. The main
idea is to expand the vertex 
function $\Gamma^{(2)}$, given by (\ref{gamma2}), in such  a way that we can get rid of the momentum dependence in the quark energies $\epsilon(p)$. Then it
will be possible to use the quasipotential equation (\ref{quas}) in order
to perform one of the integrations in the current matrix element
(\ref{mxet}).  

For the heavy-to-heavy $\Lambda_b\to\Lambda_c$ weak transitions, using
the fact that both the initial and final  baryons contain heavy
quarks, one can expand the decay matrix elements in inverse powers of
the heavy quark masses. Such an expansion was performed in our model
up to subleading  order in Ref.~\cite{hbardecay}. It was found that
all heavy quark symmetry relations are satisfied in our model. However
the $1/m_Q$  corrections turn out to be rather large, significantly
larger than for the heavy-to-heavy mesons transitions. This is the
consequence of the larger value of the expansion parameter
$\bar\Lambda$. Indeed in the case of baryon decays the parameter
$\bar\Lambda$ is determined by the light diquark energies
\cite{hbardecay}, while for meson decays it is determined by light
quark energies  \cite{hmdecay}. Therefore consideration of such decays
without the heavy quark expansion can significantly improve the precision of predictions.  Also such an expansion cannot be applied for the heavy-to-light $\Lambda_b\to p$ weak transitions, since the final baryon contains only light $u$ and $d$ quarks. 

It is important to take into account that  both
$\Lambda_b\to\Lambda_cl\nu_l$ and $\Lambda_b\to pl\nu_l$ decays have a
broad kinematical range. The square of the momentum transfer to the lepton pair $q^2$ varies  from 0 to $q^2_{\rm max}\approx 12$~GeV$^2$ for decays to $\Lambda_c$ and from 0 to $q^2_{\rm  max}\approx 22$~GeV$^2$ for decays to $p$. As a result the recoil momentum of the final baryon $|{\bf\Delta}|$ is almost always significantly larger than the relative quark momentum in the baryon. Thus one can neglect small
relative momentum $|{\bf p}|$ with respect to the recoil momentum $|{\bf\Delta}|$ in the energies of
quarks in energetic final baryons and replace $\epsilon_{q}(p+\Delta)\equiv\sqrt{m_{q}^2+({\bf 
p}+{\bf\Delta})^2} $ by $\epsilon_{q}(\Delta)\equiv
\sqrt{m_{q}^2+{\bf\Delta}^2}$. It is important to point out that we
keep the quark mass  in the energies $\epsilon_{q}(\Delta)$. Thus the
resulting expressions are valid both for the heavy-to-light and
heavy-to-heavy $\Lambda_b$ baryon decays. Such replacement is made in the subleading
contribution $\Gamma_\mu^{(2)}({\bf p},{\bf q})$ only and  permits us to
perform  one of the integrations  using the quasipotential
equation. As a result, the weak decay matrix elements are expressed
through the usual overlap integral of initial and final baryon wave
functions. Note that the subleading contributions are proportional to
the ratios of baryon binding energies, which are small, to the quark
energies and thus turn out to be also small numerically. Therefore we
obtain reliable expressions for the form factors in the whole
accessible kinematical range. The largest uncertainty, which turns out
to be small numerically, occurs for the heavy-to-light transitions in
the narrow region near zero recoil of the final light baryon, where
the above discussed replacement is less justified. It is important to
emphasize that we consistently take into account all relativistic
corrections including boosts of the baryon wave functions from the
rest frame to the moving one, given by Eq.~(\ref{wig}).   The obtained
expressions for the form factors are presented in the Appendix (to
simplify these expressions we set the long-range anomalous
chromomagnetic quark moment $\kappa=-1$).

It is easy to check that for the heavy-to-heavy semileptonic decays
 one can reproduce the model independent relations of the HQET \cite{iwb} by expanding the form factors
(\ref{f1})-(\ref{eq:g3s}) in inverse powers of the initial and final heavy quark masses. The resulting expressions for the leading
and subleading in $1/m_Q$ Isgur-Wise functions coincide with the ones
obtained in our previous analysis of heavy baryon decays in the framework of the heavy quark
expansion \cite{hbardecay}.   
On the other hand, for the heavy-to-light decays the following HQET
relations  \cite{prc} are also valid:
\begin{eqnarray}
  \label{eq:hl}
  &&F_1(q^2)=\xi_1^{(0)}(q^2)-\xi_2^{(0)}(q^2),\qquad
            G_1(q^2)=\xi_1^{(0)}(q^2)+\xi_2^{(0)}(q^2),\cr
&&F_2(q^2)=G_2(q^2)=2\xi_2^{(0)}(q^2),\qquad F_3(q^2)=G_3(q^2)=0.
\end{eqnarray}
They arise in the infinitely heavy quark mass limit $m_Q\to\infty$
for the initial heavy  $\Lambda_Q$ baryon only.
The other form factor relations found
in the additional limits of small and large recoil \cite{feldmann} of the final
light $\Lambda_q$ baryon in the rest frame of the decaying  heavy
$\Lambda_Q$ are also satisfied.

For numerical calculations of the form factors we use the quasipotential wave functions of the 
 $\Lambda_b$, $\Lambda_c$ and $p$ baryons obtained in their mass spectra
calculations. Note that these calculations were done without the
application of nonrelativistic $v/c$ and heavy quark $1/m_Q$
expansions. Therefore the resulting wave functions  incorporate
nonperturbatively the relativistic quark dynamics in heavy and light baryons. Our results for the masses of these baryons are in good agreement with experimental data \cite{pdg}, which we use in our calculations. 
We find that the weak decay baryon form factors can 
be approximated with good accuracy by the following expressions: 
\begin{equation}
  \label{fitff}
  F(q^2)=\frac{F(0)}{\displaystyle\left(1-\sigma_1\frac{q^2}{M^2_{\Lambda_Q}}+\sigma_2  \frac{q^4}{M_{\Lambda_Q}^4}+
      \sigma_3\frac{q^6}{M_{\Lambda_Q}^6}+
      \sigma_4\frac{q^8}{M_{\Lambda_Q}^8}\right)}.
\end{equation}
The difference of fitted form factors from the calculated ones does not exceed  0.5\%.
 
 The values  $F(0)$, $F(q^2_{\rm max})$ and $\sigma_{1,2,3,4}$ are given in 
Tables~\ref{ffLbLc},~\ref{ffLbp}. The evaluation of the theoretical
uncertainties of form factor calculations  represents an
important issue. Of course, the uncertainty of the model itself is not
known since it is not directly derived from QCD.  We can estimate the
errors only within our model. They mostly originate from the
uncertainties in the baryon wave functions and for the heavy-to-light
transitions from the subleading contribution in the low recoil
region. For example, to estimate the errors coming from the baryon wave functions we compared the form factors calculated on the basis of the complete relativistic wave functions with the corresponding ones calculated on the basis of the wave functions obtained in the heavy quark limit. As a result we find that the total error of our form factors should  
be less than 5\%.  

In Table~\ref{compbpiff} the comparison of
theoretical predictions for the form factors $f^{V,A}_{1,2,3}(0)$ is
given. Calculations in Refs.~\cite{giklsh,gikls} are based on the covariant
confined  quark model. The authors of Ref.~\cite{wkl} use the light front quark model and diquark picture, while QCD light-cone sum rules are employed in Ref.~\cite{kkmw}. Reasonable agreement between predictions of significantly different approaches for calculating baryon form factors is observed.

\begin{table}
\caption{Calculated form factors of the weak $\Lambda_b\to \Lambda_c$ transition. }
\label{ffLbLc}
\begin{ruledtabular}
\begin{tabular}{ccccccc}
& $F_1(q^2)$ & $F_2(q^2)$& $F_3(q^2)$& $G_1(q^2)$ & $G_2(q^2)$ &$G_3(q^2)$\\
\hline
$F(0)$          &0.719 &$-0.062$ & $-0.086$ & 0.520 & $-0.225$&0.113\\
$F(q^2_{\rm max})$&1.62  &$-0.304$ & $-0.218$ & 1.11& $-0.611$& 0.314\\
$\sigma_1$      &$1.46$&$2.28$& $2.11$& $1.46$ &$1.56$&  2.11\\
$\sigma_2$      &$-4.27$&$-7.98$&$-0.99$&$-3.06$&$-6.44$&$-2.49$\\
$\sigma_3$      &$29.1$&$53.6$& $17.2$& $22.3$ &$42.3$&  25.4\\
$\sigma_2$      &$-51.1$&$-87.5$&$-31.7$&$-39.9$&$-73.7$&$-45.3$\\
\end{tabular}
\end{ruledtabular}
\end{table}

\begin{table}
\caption{Calculated form factors of the weak $\Lambda_b\to p$ transition. }
\label{ffLbp}
\begin{ruledtabular}
\begin{tabular}{ccccccc}
& $F_1(q^2)$ & $F_2(q^2)$& $F_3(q^2)$& $G_1(q^2)$ & $G_2(q^2)$ &$G_3(q^2)$\\
\hline
$F(0)$          &0.227 &$-0.021$ & $-0.013$ & 0.196 & $-0.076$&0.013\\
$F(q^2_{\rm max})$&1.50  &$-0.463$ & $-0.144$ & 0.905& $-0.748$& 0.283\\
$\sigma_1$      &$0.805$&$2.80$& $1.72$& $0.712$ &$1.06$&  1.87\\
$\sigma_2$      &$-3.06$&$2.14$&$-2.07$&$-2.23$&$-4.49$&$-2.64$\\
$\sigma_3$      &$4.90$&$0.60$& $6.18$& $2.85$ &$10.2$&  8.45\\
$\sigma_2$      &$-1.96$&$-1.07$&$-3.37$&$-0.71$&$-6.08$&$-5.24$\\
\end{tabular}
\end{ruledtabular}
\end{table}

\begin{table}
\caption{Comparison of theoretical predictions for the form factors of 
  weak baryon decays at maximum
  recoil point $q^2=0$.  }
\label{compbpiff}
\begin{ruledtabular}
\begin{tabular}{ccccccc}
&$f^V_1(0)$&$f^V_2(0)$&$f^V_3(0)$&$f^A_1(0)$&$f^A_2(0)$&$f^A_3(0)$\\
\hline
$\Lambda_b\to\Lambda_c$\\
this paper& 0.526& 0.137& 0.075& 0.505& $-0.027$& $-0.252$\\
\cite{giklsh}&0.549&0.110& $-0.023$& 0.542& 0.018&$-0.123$\\
\cite{wkl} &0.5057& 0.0994& &0.5009 &0.0089\\
$\Lambda_b\to p$\\
this paper& 0.169& 0.050& 0.029& 0.196& $-0.0002$& $-0.076$\\
\cite{gikls}&0.080&0.036& $-0.005$& 0.077& $-0.001$&$-0.046$\\
\cite{kkmw}&$0.12^{+0.03}_{-0.04}$&$0.047^{+0.015}_{-0.013}$&
&$0.14^{+0.03}_{-0.03}$ &$-0.016^{+0.007}_{-0.005}$&\\
\cite{wkl} &0.1131& 0.0356& &0.112 &0.0097\\
\end{tabular}
\end{ruledtabular}
\end{table}

\begin{figure}
\centering
  \includegraphics[width=8cm]{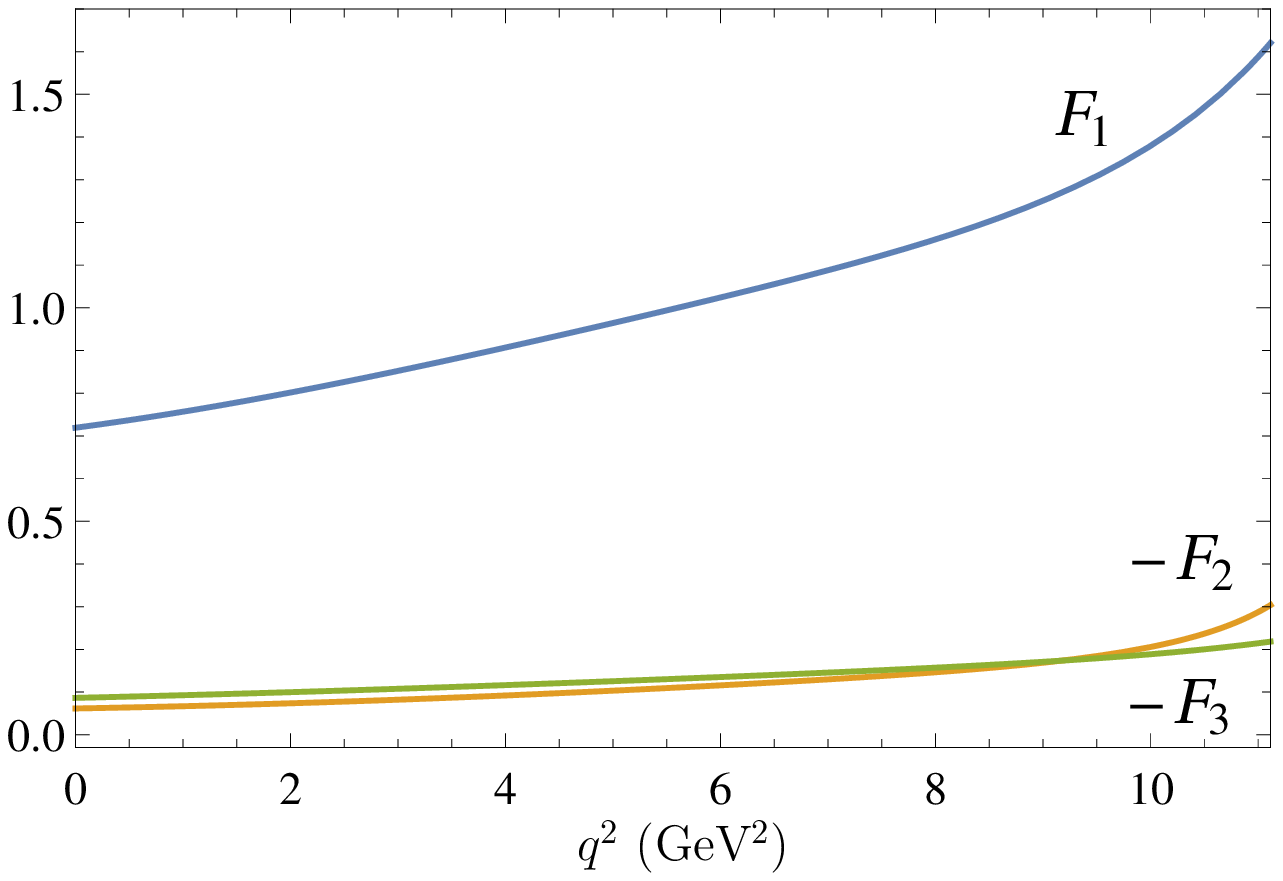}\ \
 \ \includegraphics[width=8cm]{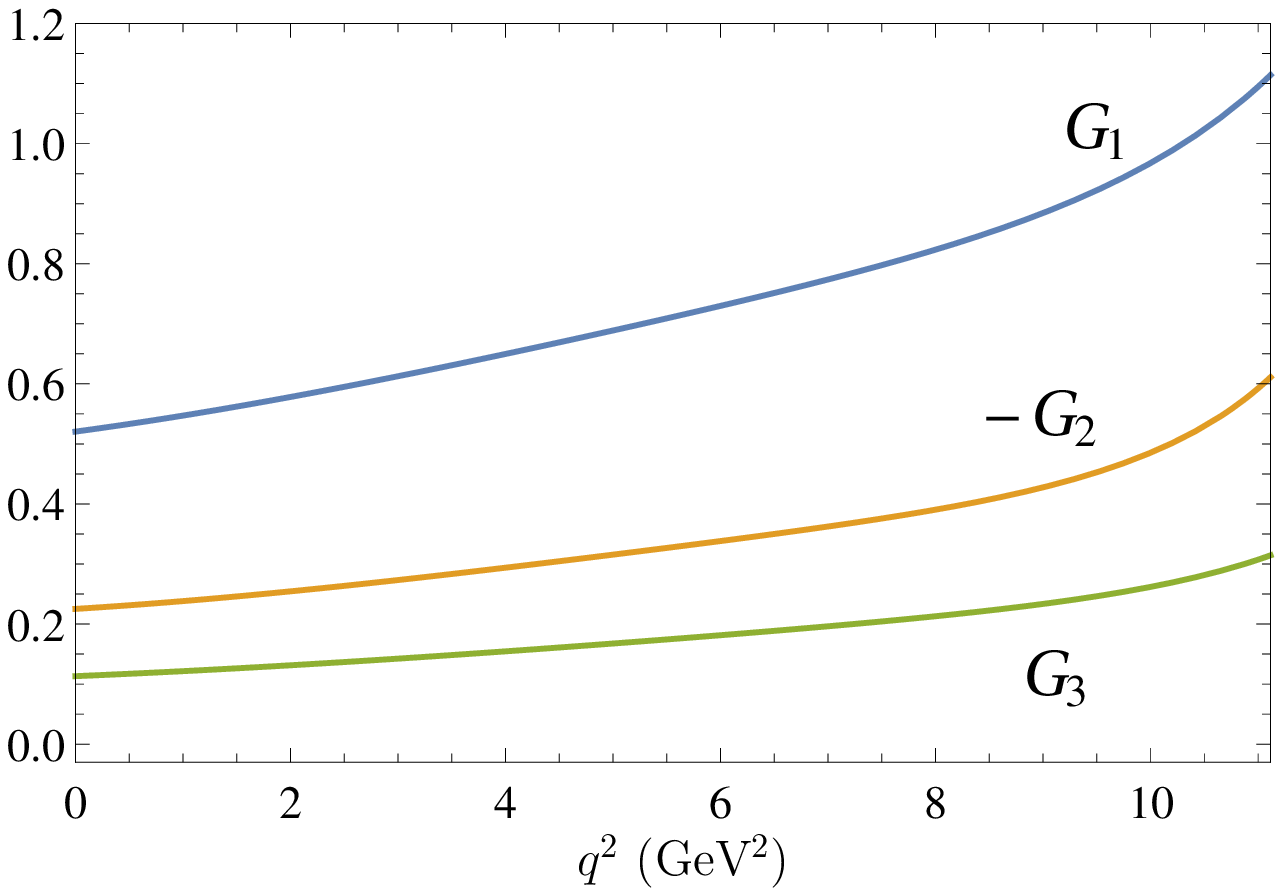}\\
\caption{Form factors of the weak $\Lambda_b\to \Lambda_c$ transition.    } 
\label{fig:ffLbLs}
\end{figure}

\begin{figure}
\centering
  \includegraphics[width=8cm]{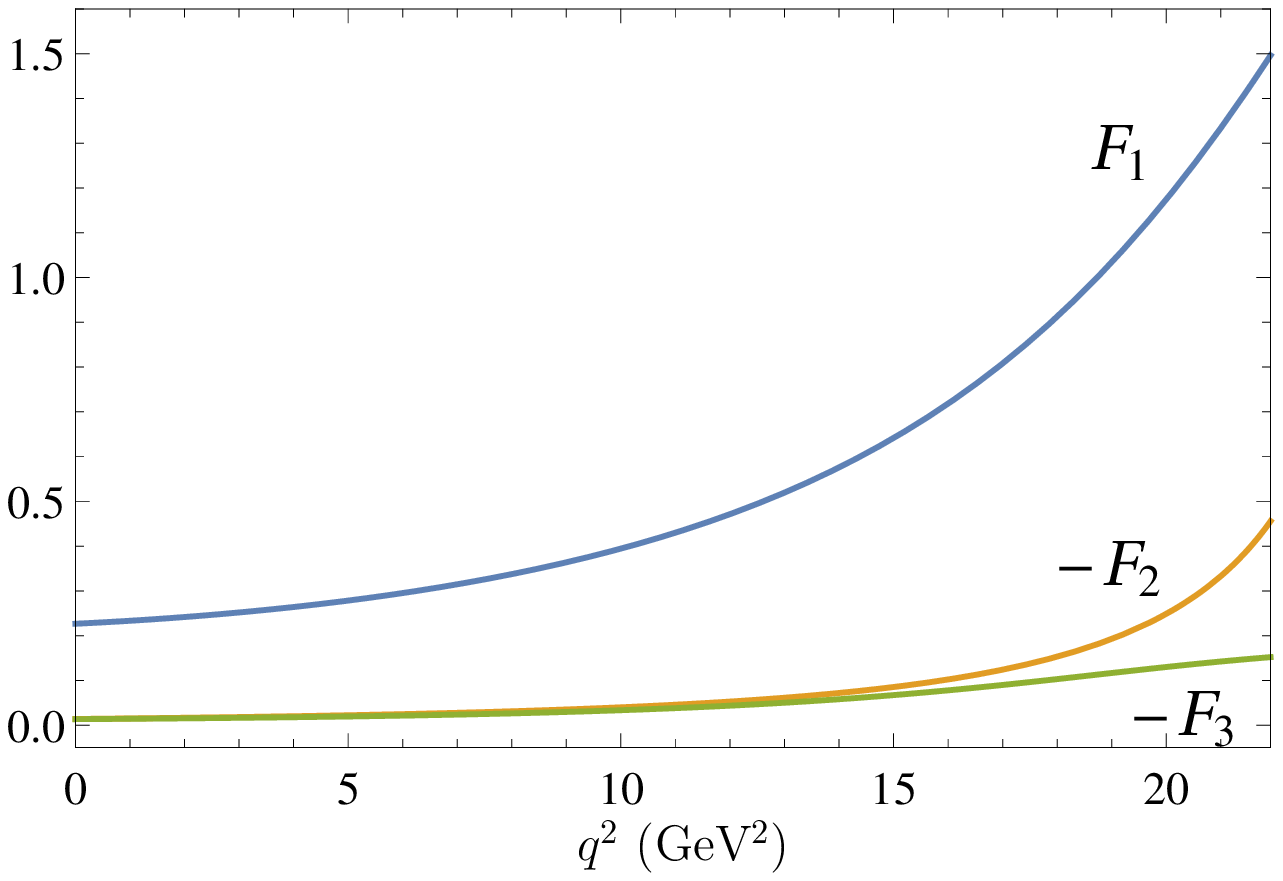}\ \
 \ \includegraphics[width=8cm]{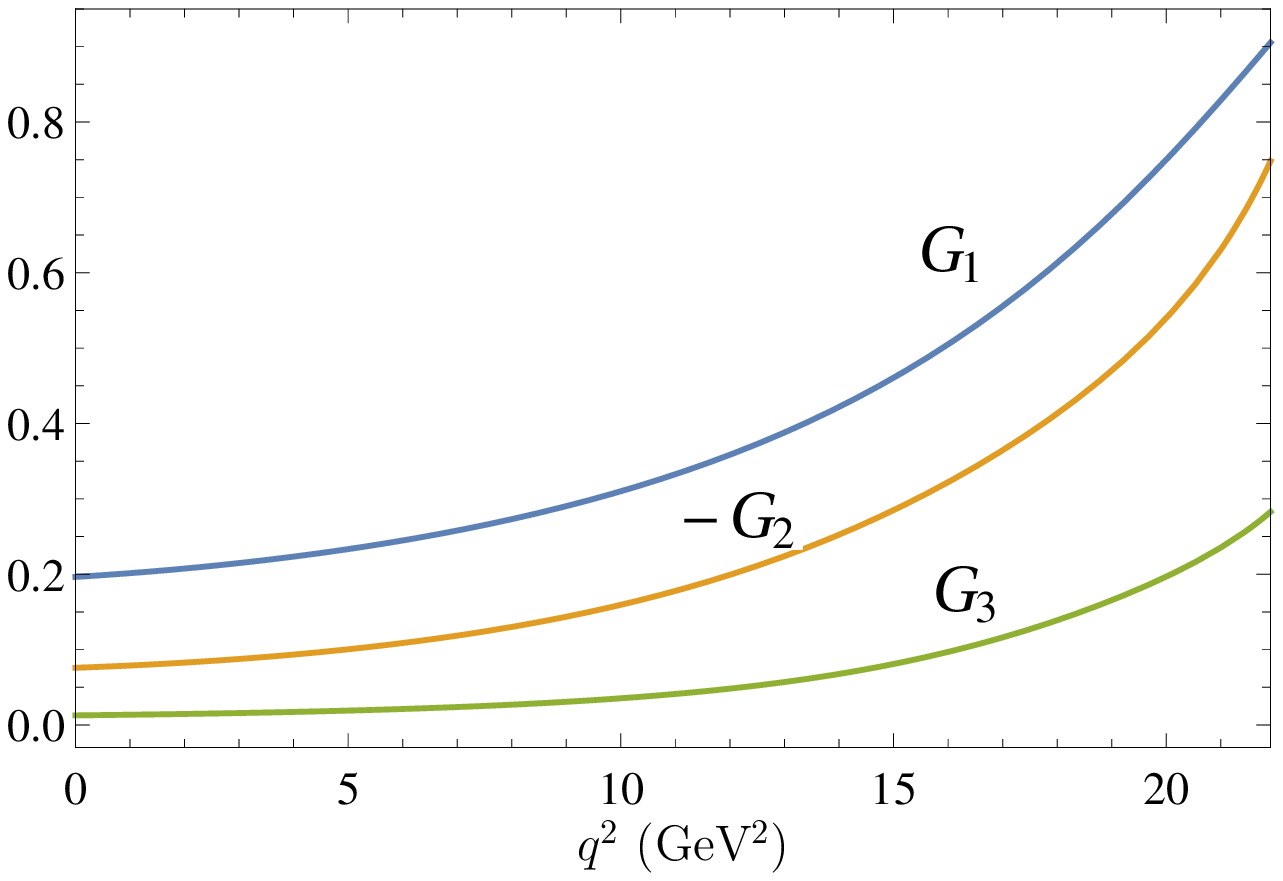}\\
\caption{Form factors of the weak $\Lambda_b\to p$ transition.    } 
\label{fig:ffLbp}
\end{figure}

We plot the baryon decay form factors in Figs.~\ref{fig:ffLbLs} and
\ref{fig:ffLbp}. The comparison of the $\Lambda_b \to\Lambda_c$ form
factor plots in Fig.~\ref{fig:ffLbLs} with our previous calculation
within heavy quark expansion presented in Fig. 5 of
Ref.~\cite{hbardecay} indicate the general consistency, but the
values of  the form factors $|F_1(q^2_{\rm max})|$, $|G_{1,2}(q^2_{\rm
  max})|$ are somewhat larger and these form factors increase with
$q^2$ more rapidly in the present unexpanded in $1/m_Q$
consideration. This observation confirms our expectations of the
importance of the nonperturbative in $1/m_Q$ treatment of the
semileptonic heavy-to-heavy baryon form factors.

\section{Heavy-to-heavy and heavy-to-light semileptonic $\Lambda_b$ baryon decays}

Now we can use the baryon form factors found in the previous section for the calculation of the $\Lambda_b$ semileptonic decay rates. For obtaining the corresponding expressions for the decay rates in terms of form factors it is convenient to use  the helicity formalism \cite{kk}. 

The helicity amplitudes are expressed in terms of the baryon form factors
\cite{kk} as
\begin{eqnarray}
  \label{eq:ha}
  H^{V,A}_{+1/2,\, 0}&=&\frac1{\sqrt{q^2}}{\sqrt{2M_{\Lambda_Q}M_{\Lambda_q}(w\mp 1)}}
[(M_{\Lambda_Q} \pm M_{\Lambda_q}){\cal F}^{V,A}_1(w) \pm M_{\Lambda_q}
(w\pm 1){\cal F}^{V,A}_2(w)\cr
&& \pm M_{\Lambda_{Q}} (w\pm 1){\cal F}^{V,A}_3(w)],\cr
 H^{V,A}_{+1/2,\, 1}&=&-2\sqrt{M_{\Lambda_Q}M_{\Lambda_q}(w\mp 1)}
 {\cal F}^{V,A}_1(w),\cr
 H^{V,A}_{+1/2,\, t}&=&\frac1{\sqrt{q^2}}{\sqrt{2M_{\Lambda_Q}M_{\Lambda_q}(w\pm 1)}}
[(M_{\Lambda_Q} \mp M_{\Lambda_q}){\cal F}^{V,A}_1(w) \pm(M_{\Lambda_Q}- M_{\Lambda_q}w
){\cal F}^{V,A}_2(w)\cr
&& \pm (M_{\Lambda_{Q}} w- M_{\Lambda_q}){\cal F}^{V,A}_3(w)],
\end{eqnarray}
where  
$$w=\frac{M_{\Lambda_Q}^2+M_{\Lambda_{q}}^2-q^2}
{2M_{\Lambda_Q}M_{\Lambda_{q}}},$$ 
the upper(lower)  sign corresponds  to $V(A)$ and ${\cal F}^V_i\equiv F_i$,
${\cal F}^A_i\equiv G_i$ ($i=1,2,3$). $H^{V,A}_{\lambda',\,
  \lambda_W}$ are the helicity  
amplitudes for weak transitions induced by vector ($V$) and axial
vector ($A$) currents, where $\lambda'$ and $\lambda_W$ are the
helicities of the final baryon and the virtual $W$-boson, respectively. 
The amplitudes for negative values of the helicities can be obtained
using the relation
$$H^{V,A}_{-\lambda',\,-\lambda_W}=\pm H^{V,A}_{\lambda',\, \lambda_W}.$$
The total helicity amplitude for the
$V-A$ current is then given by
\begin{equation}
\label{ha}
H_{\lambda',\, \lambda_W}=H^{V}_{\lambda',\, \lambda_W}
-H^{A}_{\lambda',\, \lambda_W}.
\end{equation}

Following Ref.~\cite{giklsh} we write the twofold angular distribution
  for the decay $\Lambda_Q\to \Lambda_q W^-(\to \ell^-\bar\nu_\ell)$
\begin{equation}
\label{eq:ddgamma}
  \frac{d\Gamma(\Lambda_Q\to \Lambda_q\ell\bar\nu_\ell)}{dq^2d\cos\theta}=\frac{G_F^2}{(2\pi)^3}
  |V_{qQ}|^2
  \frac{\lambda^{1/2}(q^2-m_\ell^2)^2}{48M_{\Lambda_Q}^3q^2}W(\theta,q^2),
\end{equation}
with
\begin{eqnarray}
  \label{eq:w}
  W(\theta,q^2)&=&\frac38\Biggl\{(1+\cos^2\theta){\cal
               H}_U(q^2)-2\cos\theta{\cal H}_P(q^2)+ 2\sin^2\theta{\cal H}_L(q^2)\cr
&&+\frac{m_\ell^2}{q^2}\left(2{\cal H}_s(q^2)+\sin^2\theta{\cal H}_U(q^2)+2\cos^2\theta
 {\cal H}_l(q^2)-4\cos\theta{\cal H}_{SL}(q^2)\right)\Biggl\}.
\end{eqnarray}
Here $G_F$ is the Fermi constant, $V_{qQ}$ is the CKM matrix element,
 $\lambda\equiv
\lambda(M_{\Lambda_Q}^2,M_{\Lambda_q}^2,q^2)=M_{\Lambda_Q}^4+M_{\Lambda_q}^4+q^4-2(M_{\Lambda_Q}^2M_{\Lambda_q}^2+M_{\Lambda_q}^2q^2+M_{\Lambda_Q}^2q^2)$,
and $m_\ell$ is the lepton mass ($\ell=e,\mu,\tau$). $\theta$ is the
angle between the lepton $\ell$ and $W$ momenta. 

The relevant parity conserving helicity structures are expressed in
terms of the total helicity amplitudes (\ref{ha}) by
\begin{eqnarray}
  \label{eq:hhc}
  {\cal H}_U(q^2)&=&|H_{+1/2,+1}|^2+|H_{-1/2,-1}|^2,\cr
{\cal H}_L(q^2)&=&|H_{+1/2,0}|^2+|H_{-1/2,0}|^2,\cr
{\cal H}_S(q^2)&=&|H_{+1/2,t}|^2+|H_{-1/2,t}|^2,\cr
{\cal H}_{SL}(q^2)&=&{\rm Re}(H_{+1/2,0}H_{+1/2,t}^\dag+H_{-1/2,0}H_{-1/2,t}^\dag),
\end{eqnarray}
and the parity violating helicity structures by
\begin{eqnarray}
  \label{eq:hhv}
  {\cal H}_P(q^2)&=&|H_{+1/2,+1}|^2-|H_{-1/2,-1}|^2,\cr
{\cal H}_{L_P}(q^2)&=&|H_{+1/2,0}|^2-|H_{-1/2,0}|^2,\cr
{\cal H}_{S_P}(q^2)&=&|H_{+1/2,t}|^2-|H_{-1/2,t}|^2.
\end{eqnarray}

The differential decay rate is obtained by integrating (\ref{eq:ddgamma}) over $\cos\theta$ \cite{gikls}
\begin{equation}
  \label{eq:dgamma}
  \frac{d\Gamma(\Lambda_Q\to \Lambda_q\ell\bar\nu_\ell)}{dq^2}=\frac{G_F^2}{(2\pi)^3}
  |V_{qQ}|^2
  \frac{\lambda^{1/2}(q^2-m_\ell^2)^2}{48M_{\Lambda_Q}^3q^2}{\cal H}_{tot}(q^2),
\end{equation}
where
\begin{equation}
 \label{eq:hh}
 {\cal H}_{tot}(q^2)=[{\cal H}_U(q^2)+{\cal H}_L(q^2)] \left(1+\frac{m_\ell^2}{2q^2}\right)+\frac{3m_\ell^2}{2q^2}{\cal H}_S(q^2) .
\end{equation}

\begin{figure}
  \centering
 \includegraphics[width=8cm]{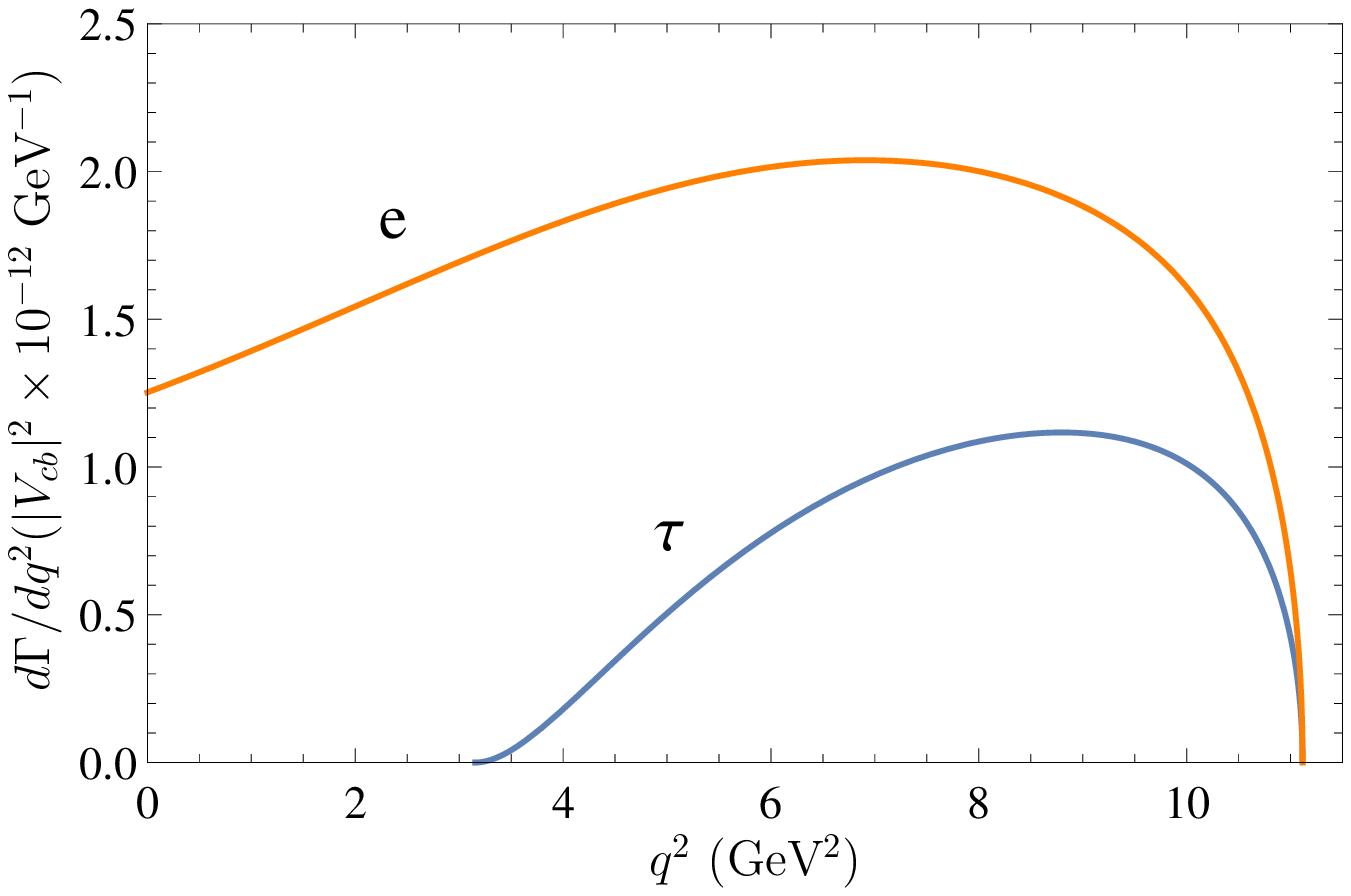}\ \
 \  \includegraphics[width=8cm]{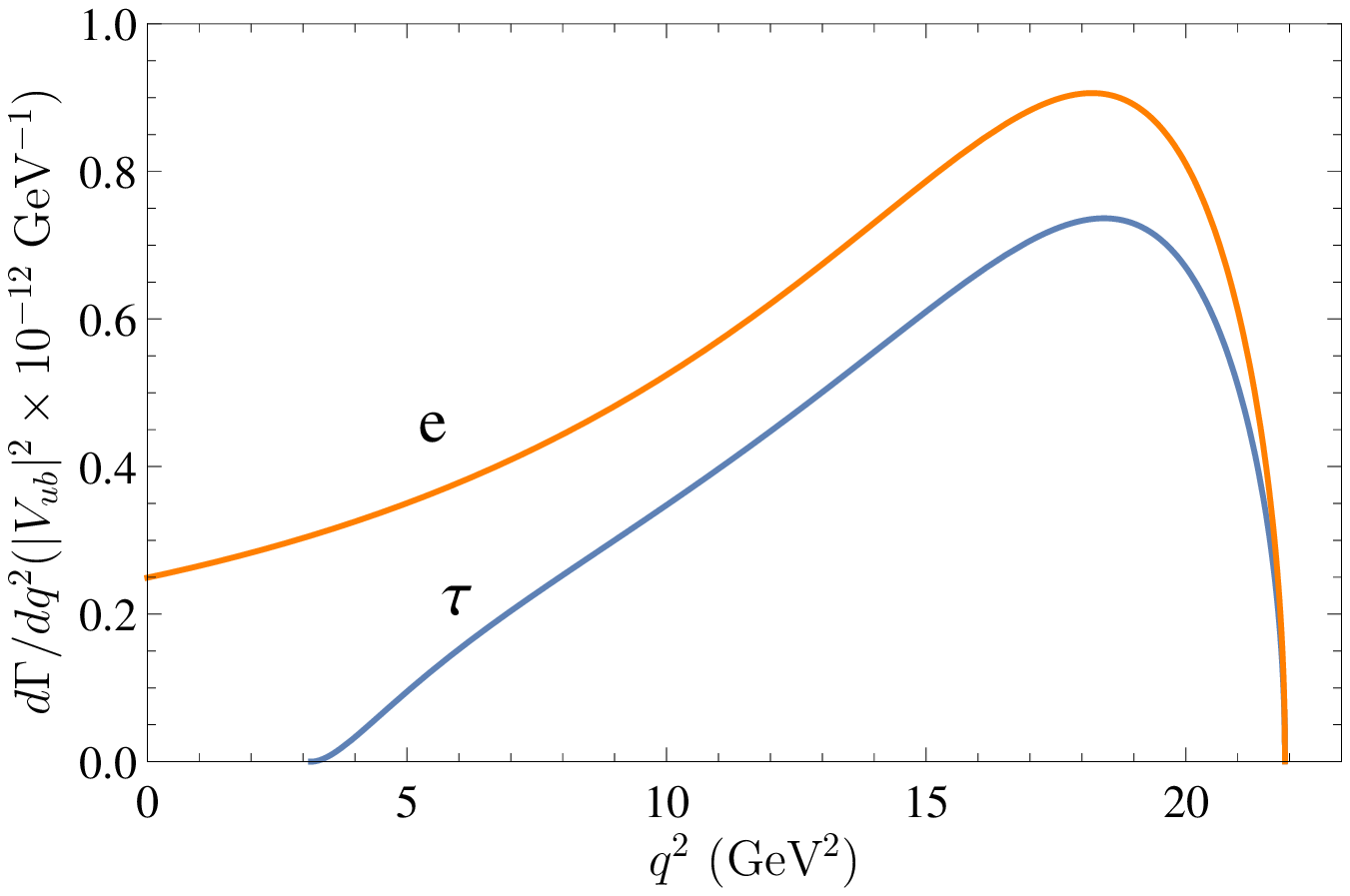}

  \caption{Predictions for the differential decay rates  of the
    $\Lambda_b\to \Lambda_c\ell\nu_\ell$ (left) and $\Lambda_b\to p\ell\nu_\ell$ (right)
    semileptonic decays. }
  \label{fig:brLb}
\end{figure}

Substituting in these expressions the  baryon form factors calculated in our model in the previous section we obtain corresponding semileptonic differential decay rates which are plotted in Fig.~\ref{fig:brLb}.

\begin{figure}
  \centering
 \includegraphics[width=8cm]{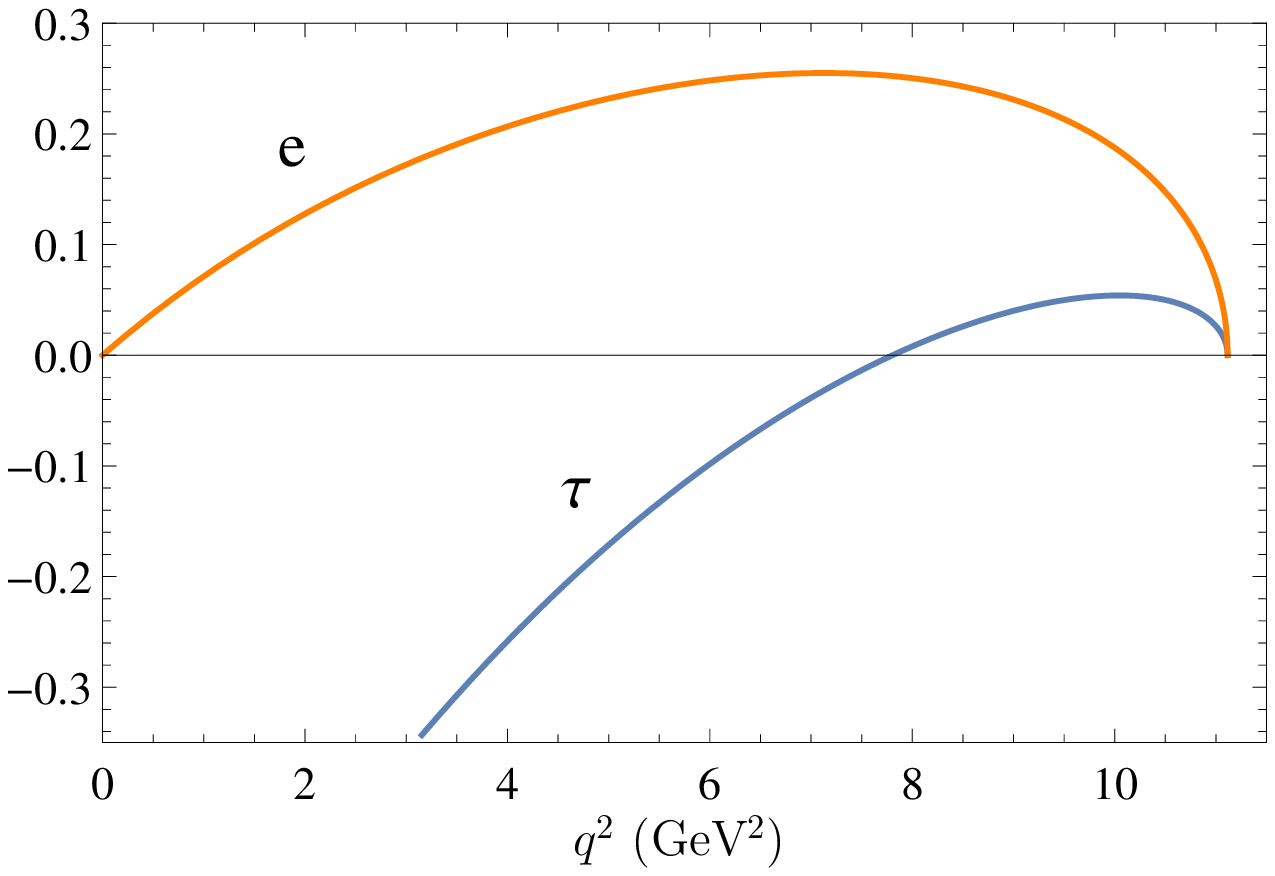}\ \
 \  \includegraphics[width=8cm]{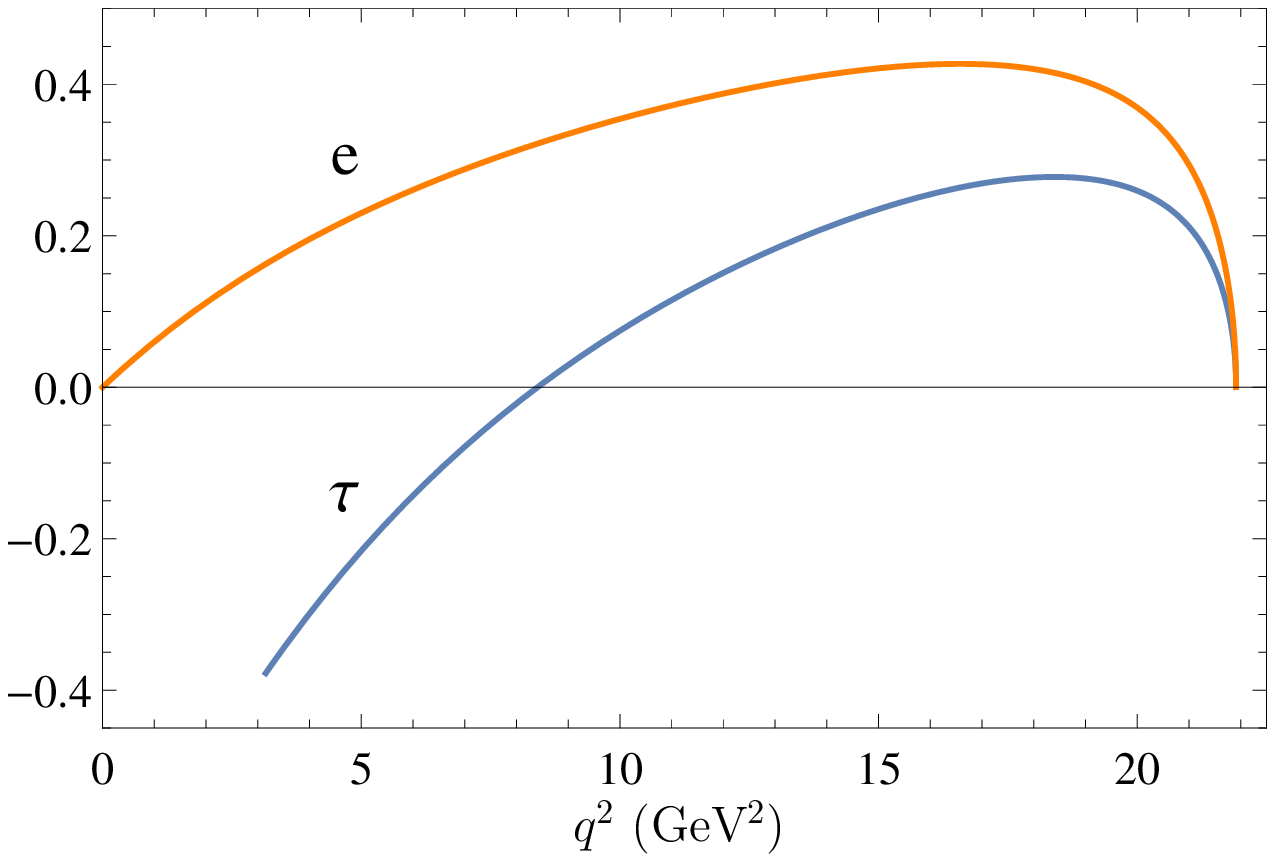}

  \caption{Predictions for the forward-backward asymmetries $A_{FB}(q^2)$ in the
    $\Lambda_b\to \Lambda_c\ell^-\nu_\ell$ (left) and $\Lambda_b\to p\ell^-\nu_\ell$ (right)
    semileptonic decays. }
  \label{fig:afbLb}
\end{figure}

\begin{figure}
  \centering
 \includegraphics[width=8cm]{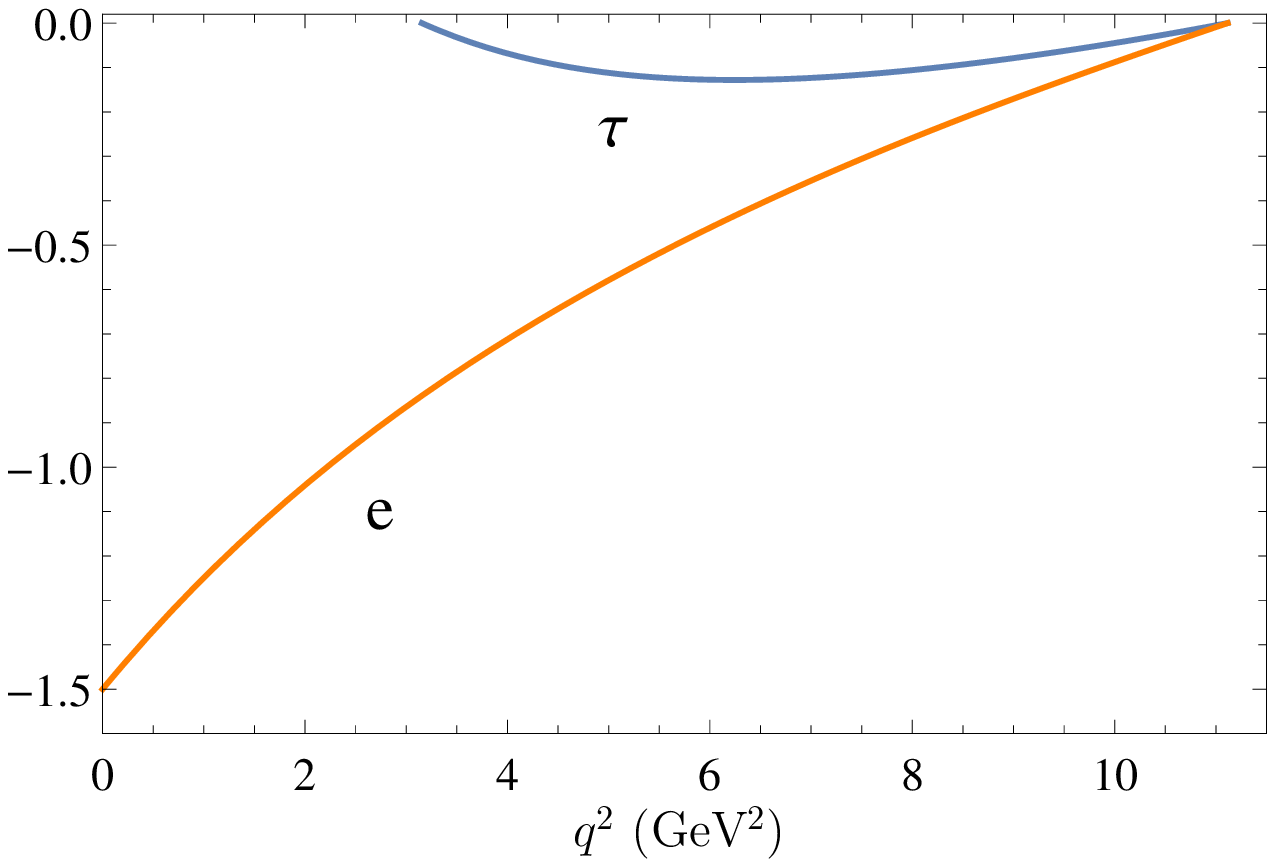}\ \
 \  \includegraphics[width=8cm]{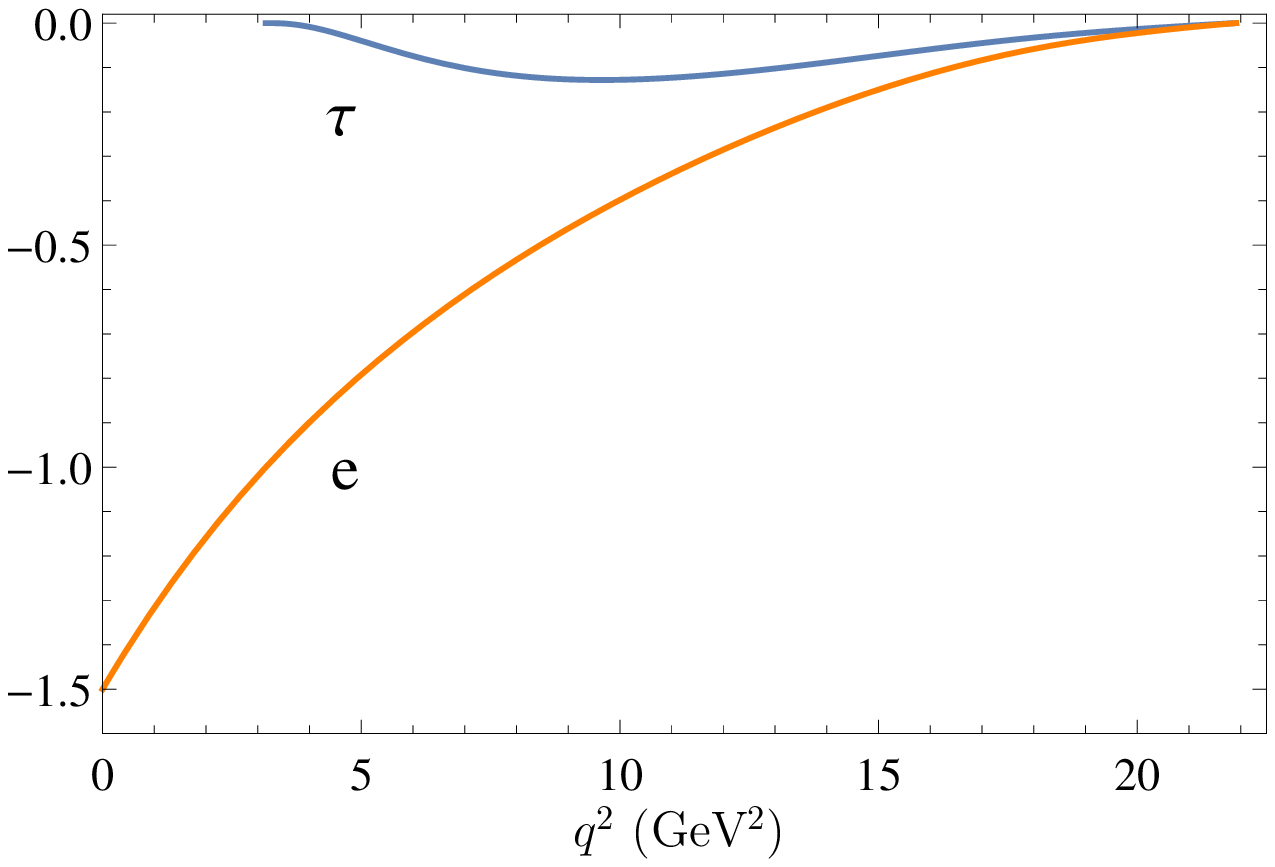}

  \caption{Predictions for the convexity parameter  $C_F(q^2)$ in the
    $\Lambda_b\to \Lambda_c\ell\nu_\ell$ (left) and $\Lambda_b\to p\ell\nu_\ell$ (right)
    semileptonic decays. }
  \label{fig:cfLb}
\end{figure}

\begin{figure}
  \centering
 \includegraphics[width=8cm]{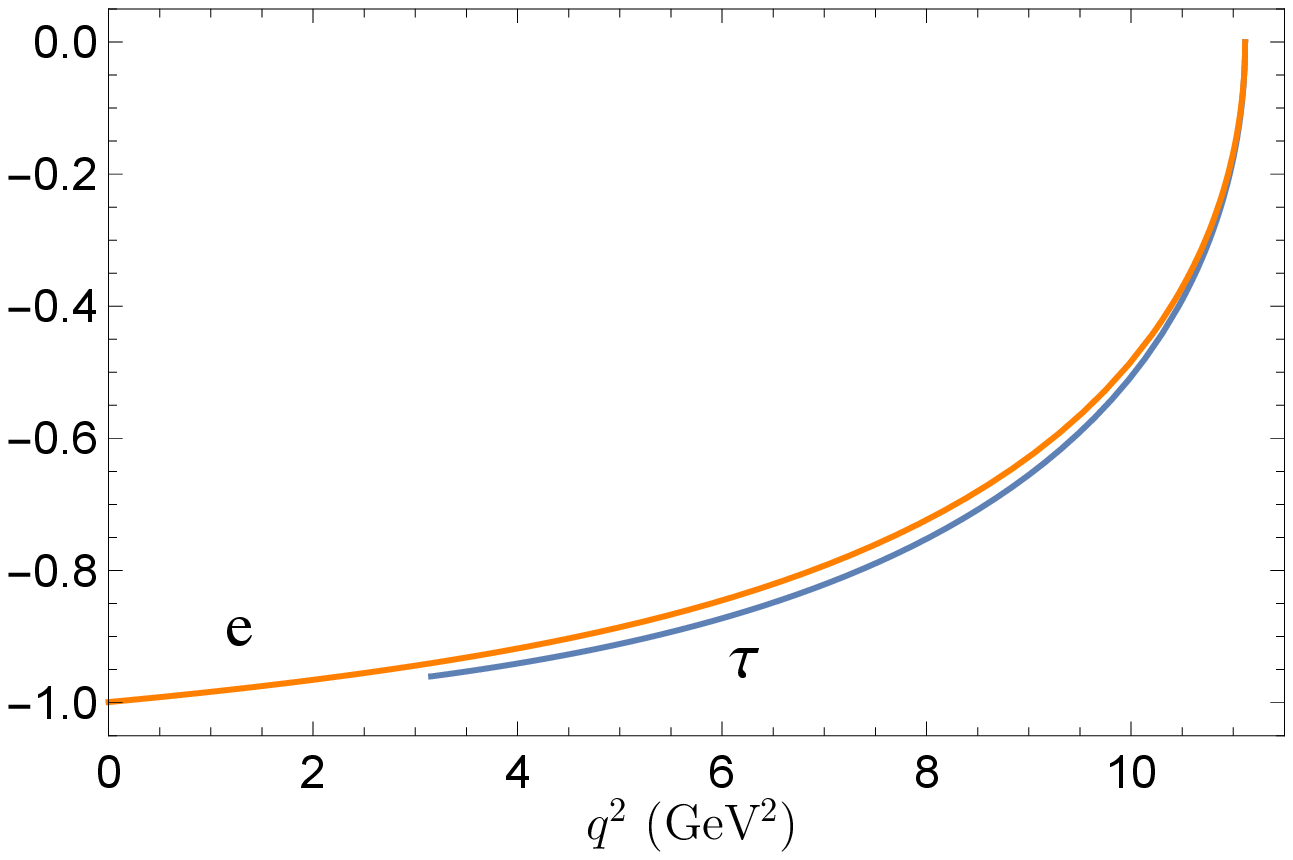}\ \
 \  \includegraphics[width=8cm]{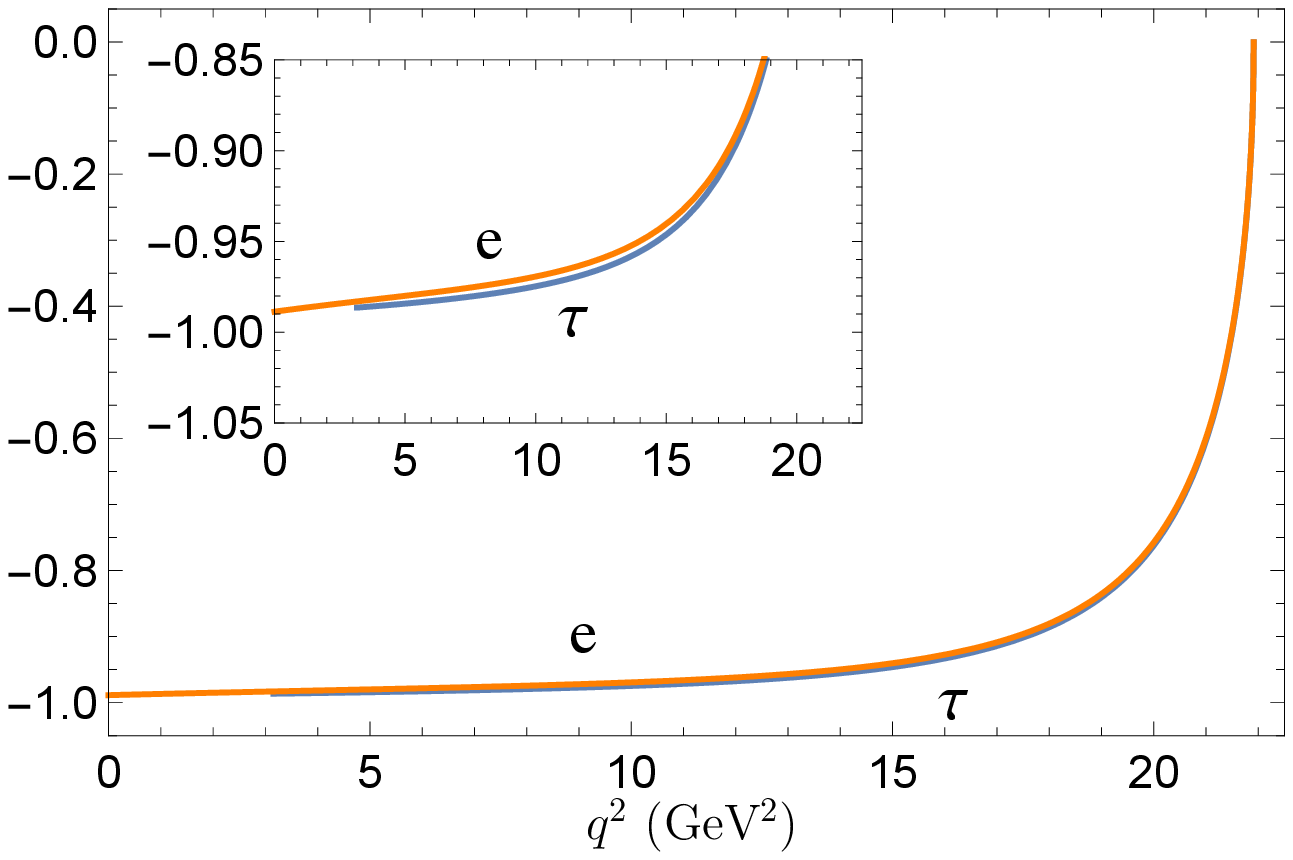}

  \caption{Predictions for the longitudinal polarization $P_L(q^2)$ of the final baryon  in the
    $\Lambda_b\to \Lambda_c\ell\nu_\ell$ (left) and $\Lambda_b\to p\ell\nu_\ell$ (right)
    semileptonic decays. }
  \label{fig:PLLb}
\end{figure}

Many important observables can also be expressed in terms of the
helicity amplitudes \cite{gikls}.
The forward-backward asymmetry of the charged lepton is the term linear in
$\cos\theta$  in the distribution (\ref{eq:w}) given by 
\begin{equation}
  \label{eq:afb}
  A_{FB}(q^2)=\frac{\frac{d\Gamma}{dq^2}({\rm forward})-\frac{d\Gamma}{dq^2}({\rm backward})}{\frac{d\Gamma}{dq^2}}
=-\frac34\frac{{\cal H}_P(q^2)+2\frac{m_\ell^2}{q^2}{\cal H}_{SL}(q^2)}{{\cal H}_{tot}(q^2)}.\qquad
\end{equation}
The term quadratic in $\cos\theta$  in the distribution (\ref{eq:w}) is the convexity parameter defined
by 
\begin{equation}
  \label{eq:cf}
  C_F(q^2)=\frac1{{\cal H}_{tot}(q^2)}\frac{d^2W(\theta,q^2)}{d(\cos\theta)^2}=\frac34\left(1-\frac{m_\ell^2}{q^2}\right)\frac{{\cal H}_U(q^2)-2{\cal H}_L(q^2)}{{\cal H}_{tot}(q^2)}.
\end{equation}
The longitudinal polarization of the final baryon $\Lambda_q$ reads as
\begin{equation}
P_L(q^2)=\frac{[{\cal H}_P(q^2)+{\cal H}_{L_P}(q^2)]\left(1+\frac{m_\ell^2}{2q^2}\right)+3 \frac{m_\ell^2}{2q^2}{\cal H}_{S_P}(q^2)}{{\cal H}_{tot}(q^2)}.
\end{equation}
The plots for these observables are given in Figs.~\ref{fig:afbLb}-\ref{fig:PLLb} for both heavy-to-heavy $\Lambda_b\to\Lambda_c$ and heavy-to-light $\Lambda_b\to p$ semileptonic decays.

Integrating the differential decay rate (\ref{eq:dgamma}) we get our
predictions for the total decay rates and branching ratios which are
given in Table~\ref{pred}. We estimate the errors of our calculations
of the decay rates and branching fractions divided by the square of
the corresponding CKM matrix element  $|V_{qQ}|^2$, to be about
10\%. For absolute values we use the following CKM values
$|V_{cb}|=(3.90\pm0.15)\times 10^{-2}$, 
$|V_{ub}|=(4.05\pm0.20)\times10^{-3}$ extracted from our previous
analysis of the heavy $B$ and $B_s$ meson decays \cite{slbdecay}. In
this table we also give our predictions for the average values of the
forward-backward asymmetry of the charged lepton $\langle
A_{FB}\rangle$, the convexity parameter $\langle C_F\rangle$ and the
longitudinal polarization of the final baryon $\langle P_L\rangle$
which are calculated by separately integrating the numerators and
denominators over $q^2$. Note that these quantities are less sensitive
to the uncertainties in the form factor calculations since the errors
partially cancel in  the ratios of the helicity structures. We find
the uncertainties of our predictions for them to be about 3-4\%.   

\begin{table}
\caption{Predictions for baryon decay rates, branching fractions and
  asymmetry parameters. }
\label{pred}
\begin{ruledtabular}
\begin{tabular}{cccccccc}
Decay& $\Gamma$ (ns$^{-1}$) & $\Gamma/|V_{qQ}|^2$ (ps$^{-1}$)& $Br$ (\%)&$Br/|V_{qQ}|^2$
  & $\langle A_{FB}\rangle$ &$\langle C_F\rangle$& $\langle P_L\rangle$\\
\hline
$\Lambda_b\to\Lambda_ce\nu_e$ &44.2&29.1 & 6.48 & 42.6 & $0.195$&$-0.57$&$-0.80$\\
$\Lambda_b\to\Lambda_c\mu\nu_\mu$ &44.1&29.0 & 6.46 & 42.5 & $0.189$&$-0.55$&$-0.80$\\$\Lambda_b\to\Lambda_c\tau\nu_\tau$&13.9  &9.11 & 2.03 & 13.4& $-0.021$& $-0.09$&$-0.71$\\
$\Lambda_b\to pe\nu_e$ &0.306&18.7 & 0.045 & 27.4 & $0.346$&$-0.32$&$-0.91$\\
$\Lambda_b\to p\mu\nu_\mu$ &0.306&18.7 & 0.045 & 27.4 & $0.344$&$-0.32$&$-0.91$\\$\Lambda_b\to p\tau\nu_\tau$&0.199  &12.1 & 0.029 & 17.8& $-0.185$& $-0.09$&$-0.89$\\
\end{tabular}
\end{ruledtabular}
\end{table}

We compare our predictions with the results of other theoretical
approaches \cite{giklsh,gikls,prc,dutta,wkl,kkmw,lat} and available
experimental data \cite{pdg} in Table~\ref{comp}.~\footnote{We limit
  our comparison to the recent results only. References to previous
  predictions and comparison with them can be found, e.g., in
  Refs.~\cite{giklsh,gikls}}  The most comprehensive results for
different decay parameters were previously obtained in the covariant
confined  quark model (CCQ) \cite{giklsh,gikls}, with which we find
the general agreement. The semirelativistic quark model is used in
Ref.~\cite{prc}, while effective Lagrangian approach with form factors
calculated on the lattice \cite{lat} is employed in
Ref.~\cite{dutta}. The authors of Refs.~\cite{wkl,kkmw} made
calculations in the light-front quark model and in QCD light-cone sum
rules, respectively. The only experimental data are available for the
branching ratio of the $\Lambda_b\to\Lambda_c^+l^-\bar\nu_l$ decay ($l=e,\mu$). All theoretical predictions agree well with data within error bars. However, note that lattice calculations \cite{lat} give somewhat lower predictions for the branching ratios normalized by the square of the corresponding CKM matrix element for $\Lambda_b\to\Lambda_c$ transitions but give higher results for $\Lambda_b\to p$ transitions than other approaches.

\begin{table}
\caption{Comparison of theoretical predictions for the $\Lambda_b$  semileptonic decay
  parameters with available experimental data. }
\label{comp}
\begin{ruledtabular}
\begin{tabular}{ccccccccc}
Parameter& this paper & \cite{giklsh,gikls} & \cite{prc}
  & \cite{dutta}&\cite{wkl}&\cite{kkmw}&\cite{lat} &Exp. \cite{pdg}\\
\hline
$\Lambda_b\to\Lambda_cl\nu_l$\\
 $\Gamma$ (ns$^{-1}$)&44.2& &53.9\\
$\Gamma/|V_{cb}|^2$ (ps$^{-1}$)&29.1& & &&& &$21.5\pm0.8\pm1.1$&\\
$Br$ (\%)& 6.48& 6.9& &4.83&6.3& &&$6.2^{+1.4}_{-1.2}$\\
$\langle A_{FB}\rangle$ &$0.195$&0.18\\
$\langle C_F\rangle$& $-0.57$& $-0.63$\\
$\langle P_L\rangle$& $-0.80$& $-0.82$\\
\hline
$\Lambda_b\to\Lambda_c\tau\nu_\tau$\\
 $\Gamma$ (ns$^{-1}$)&13.9&& 20.9\\
$\Gamma/|V_{cb}|^2$ (ps$^{-1}$)&9.11& & &&& &$7.15\pm0.15\pm0.27$&\\
$Br$ (\%)& 2.03& 2.0& &1.63& &\\
$\langle A_{FB}\rangle$ &$-0.021$&$-0.0385$\\
$\langle C_F\rangle$& $-0.09$& $-0.10$\\
$\langle P_L\rangle$& $-0.71$& $-0.72$\\
\hline
$\Lambda_b\to pl\nu_l$\\
$\Gamma/|V_{ub}|^2$ (ps$^{-1}$)&18.7&13.3 &7.55 && &&$25.7\pm2.6\pm4.6$&\\
$Br$ ($10^{-4}$)& 4.5& 2.9& &3.89&2.54 & $4.0^{+2.3}_{-2.0}$\\
$\langle A_{FB}\rangle$ &$0.346$&$0.388$\\
\hline
$\Lambda_b\to p\tau\nu_\tau$\\
$\Gamma/|V_{ub}|^2$ (ps$^{-1}$)&12.1&9.6 &6.55 && &&$17.7\pm1.3\pm1.6$&\\
$Br$ ($10^{-4}$)& 2.9& 2.1& &2.75& &\\
$\langle A_{FB}\rangle$ &$0.185$&$0.220$\\
\end{tabular}
\end{ruledtabular}
\end{table}

At present the tension between predictions of the Standard Model and
experimental data in the $B$ meson  sector is observed for the ratio of
branching ratios of semileptonic $B$ decays to $D^{(*)}$ mesons involving 
$\tau$ and a muon or electron \cite{pdg,belle}. Therefore it is very
important to search for the similar decays in the baryon sector. We can define the following ratios of the $\Lambda_b$ baryon branching fractions
\begin{eqnarray}
  \label{eq:ratio}
  R_{\Lambda_c}&=&\frac{Br(\Lambda_b\to\Lambda_c\tau\nu_\tau)}{Br(\Lambda_b\to\Lambda_c
  l\nu_l)},\cr
R_{p}&=&\frac{Br(\Lambda_b\to p\tau\nu_\tau)}{Br(\Lambda_b\to p
  l\nu_l)}.
  \end{eqnarray}
  Our predictions for these ratios are given in Table~\ref{predc} in
  comparison with calculations \cite{swd} using the QCD sum rule form
  factors and estimates \cite{dutta} based on  lattice values of weak
  decay form factors  \cite{lat}. Results of predictions  for
  $R_{\Lambda_c}$ are in good agreement, while our $R_{p}$ value is
  slightly lower than the Ref.~\cite{dutta} estimate. Note that the lattice
  determination of form factors is done in the region of small recoils
  of the final baryon $q^2\sim q^2_{\rm max}$ and then their values are
  extrapolated to the whole kinematical region, which is broad
  especially for the heavy-to-light $\Lambda_b\to p l\nu_l$ decay. In
  our model we explicitly determine the form factor $q^2$ dependence
  in the whole kinematical range without extrapolations. The possible
  contributions of new physics to these ratios are analyzed in detail
  in Refs.~\cite{swd,dutta}.

\begin{table}
\caption{Predictions for baryon decay rates, branching fractions and
  asymmetry parameters. }
\label{predc}
\begin{ruledtabular}
\begin{tabular}{cccccc}
Ratio& this paper&\cite{swd} & \cite{dutta}&\cite{lat} &Experiment
                                             (LHCb)\cite{lhcb}\\
\hline
$R_{\Lambda_c}$& 0.313&$0.29\pm0.02$ & 0.3379& $0.3318\pm0.0074\pm0.0070$\\
$R_{p}$& 0.649& &0.7071\\
$R_{\Lambda_c p}$&
                   $(0.78\pm0.08)\frac{|V_{ub}|^2}{|V_{cb}|^2}$&&0.0101&$(1.471\pm0.095\pm0.109)\frac{|V_{ub}|^2}{|V_{cb}|^2}$&$(1.00\pm0.04\pm0.08)\times 10^{-2}$\\
\end{tabular}
\end{ruledtabular}
\end{table}

Recently the LHCb collaboration \cite{lhcb} measured the ratio of the heavy-to-heavy and heavy-to-light semileptonic $\Lambda_b$ decays in the limited interval of $q^2$
   \begin{equation}
   \label{rlcp}
R_{\Lambda_c p}=\frac{\int_{15\, {\rm GeV}^2}^{q^2_{max}}\frac{d\Gamma(\Lambda_b\to p
  \mu\nu_\mu)}{dq^2}dq^2}{\int_{7\, {\rm GeV}^2}^{q^2_{max}}\frac{d\Gamma(\Lambda_b\to \Lambda_c
  \mu\nu_\mu)}{dq^2}dq^2}.
\end{equation}
Such a measurement is very important since it allows us for the first time
to extract the ratio of CKM matrix elements $|V_{ub}|/|V_{cb}|$ from
the $\Lambda_b$ baryon decays and compare it to the corresponding
ratio determined from $B$ and $B_s$ meson decays. Our prediction for
the ratio $R_{\Lambda_c p}$ in comparison with the lattice result
\cite{lat} and experimental value is given in Table~\ref{predc}. From
this table we see that our value of the coefficient in front of
$|V_{ub}|^2/|V_{cb}|^2$ is significantly lower than the lattice one. This
is the result of the above mentioned deviation of our calculation (and other quark
model calculations) from lattice predictions for normalized by the
square of CKM matrix element value for heavy-to-heavy and
heavy-to-light semileptonic $\Lambda_b$ decays. This deviation even increases in the ratio.

Comparing our result for $R_{\Lambda_c p}$ with experimental data \cite{lhcb} we can extract the ratio of the CKM matrix elements. Using our model value we find 
\begin{equation}
  \label{eq:vubvcb}
  \frac{|V_{ub}|}{|V_{cb}|}=0.113\pm0.011|_{\rm theor}\pm0.006|_{\rm exp},
\end{equation}
which is in good agreement with the experimental ratio of these matrix
elements extracted from inclusive decays \cite{pdg}
\begin{equation}
\label{rincl}
\frac{|V_{ub}|_{\rm incl}}{|V_{cb}|_{\rm incl}}=0.105\pm0.006,
\end{equation}
and  with the corresponding ratio found in our previous analysis of
exclusive semileptonic $B$ and $B_s$
meson decays [$|V_{cb}|=(3.90\pm0.15)\times 10^{-2}$,
$|V_{ub}|=(4.05\pm0.20)\times10^{-3}$] \cite{slbdecay}
\begin{equation}
\frac{|V_{ub}|}{|V_{cb}|}=0.104\pm0.012.
\end{equation}

On the other hand, the lattice value for the ratio $R_{\Lambda_c p}$ gives
\begin{equation}
\frac{|V_{ub}|}{|V_{cb}|}=0.083\pm0.004\pm0.004,
\end{equation}
which is in agreement with the corresponding CKM matrix element ratio
extracted from the comparison of lattice predictions with data on exclusive $B$ meson decays, but more than $3\sigma$ lower than the ratio extracted from inclusive $B$ meson decays (\ref{rincl}).

\section{Conclusions}

The semileptonic $\Lambda_b$ baryon decays were investigated in the
framework of the relativistic quark model based on the quasipotential
approach and quantum chromodynamics. All parameters of the model had been previously determined from the consideration of meson properties and were kept fixed in the current consideration of the baryon semileptonic decays. The relativistic quark-diquark picture was used for the calculations. The semileptonic decay form factors were obtained both for the heavy-to-heavy and heavy-to-light $\Lambda_b$ decays. They were expressed as the overlap integrals of the relativistic baryon wave functions which are known from the baryon mass spectra calculations. All form factors were obtained without employing nonrelativistic or heavy quark expansions. Their momentum transfer dependence was explicitly determined in the whole accessible kinematical range without any extrapolations. All relativistic effects including intermediate contributions of the intermediate negative energy states and relativistic transformations of the wave functions were consistently taken into account.  

The helicity formalism was employed for calculating the
$\Lambda_b\to\Lambda_cl\nu_l$, $\Lambda_b\to\Lambda_c\tau\nu_\tau$ and
$\Lambda_b\to pl\nu_l$, $\Lambda_b\to p\tau\nu_\tau$ decay rates and
branching fractions. Different additional observables such as
forward-backward asymmetry $A_{FB}$, convexity parameter $C_F$ and
final baryon polarization $P_L$ were also determined. The obtained
results were compared with previous theoretical calculations within
significantly different approaches including quark model calculations,
QCD light-cone sum rules, lattice simulations
\cite{giklsh,gikls,prc,dutta,wkl,kkmw,lat} and available experimental
data \cite{pdg,lhcb}. Most of our results agree well with the ones
obtained within the CCQ model \cite{giklsh,gikls}. 

Our value of the branching ratio of semileptonic $\Lambda_b\to\Lambda_cl\nu_l$ decay
is in good agreement with experimental measurement \cite{pdg}. From the recent LHCb data \cite{lhcb} on the ratio $R_{\Lambda_c p}$ of semileptonic $\Lambda_b\to p\mu\nu_\mu$ to $\Lambda_b\to\Lambda_c\mu\nu_\mu$ decay rates in the constrained momentum transfer $q^2$  range (\ref{rlcp}) we find the ratio  of the CKM matrix elements $|V_{ub}|/|V_{cb}|$ consistent within error bars with the corresponding ratio determined from inclusive $B$ meson decays \cite{pdg} and with the one previously obtained from the analysis of the exclusive $B$ and $B_s$ decays in our model \cite{slbdecay}.

We plan to apply the same approach within our model to the
investigation of  semileptonic $\Lambda_c$ decays, rare semileptonic
$\Lambda_b$ decays as well as nonleptonic baryon decays within the factorization approximation.  

\acknowledgements
The authors are grateful to A. Ali, D. Ebert, M. Ivanov, J. K\"orner
V. Lyubovitskij  and V. Matveev for valuable  discussions.  

\appendix*
\section{Form factors of weak $\Lambda_Q\to\Lambda_q$ transitions}

The final expressions for decay form factors are as follows (the value of the long range anomalous chromomagnetic quark moment $\kappa=-1$).

a) Vector form factors
\begin{equation}
\label{f1}
F_1(q^2)=F_1^{(1)}(q^2)+\varepsilon
F_1^{(2)S}(q^2)+(1-\varepsilon)F_1^{(2)V}(q^2),
\end{equation}

\begin{eqnarray}
\label{eq:f1}
F_1^{(1)}(q^2)&=&\int \frac{d^3p}{(2\pi)^3} \bar\Psi_F\left({\bf
    p}+\frac{2\epsilon_d}{E_F+M_F}{\bf
    \Delta}\right)\sqrt{\frac{\epsilon_Q(p)+m_Q}{2\epsilon_Q(p)}}\sqrt{\frac{\epsilon_q(p+\Delta)+m_q}{2\epsilon_q(p+\Delta)}}\cr
&&\times
\Biggl\{1+\frac{\epsilon_d}{\epsilon_q(p+\Delta)+m_q}\left[1+\frac{\epsilon_d}{\epsilon_Q(p)+m_Q}\frac{E_F-M_F}{E_F+M_F}\right]+\frac{\epsilon_d}{\epsilon_Q(p)+m_Q}\cr
&&-\frac13\frac{{\bf
    p}^2}{(\epsilon_q(p+\Delta)+m_q)(\epsilon_Q(p)+m_Q)}- \frac{\bf p
  \Delta}{E_F+M_F}\Biggl[\frac1{\epsilon_q(p+\Delta)+m_q}\cr
&&-\frac1{\epsilon_Q(p)+m_Q}+\frac{2M_F}{E_F+M_F}\frac{\epsilon_d}{(\epsilon_q(p+\Delta)+m_q)(\epsilon_Q(p)+m_Q)}\Biggr]\Biggr\}\Psi_I({\bf
  p}); 
\end{eqnarray}
\begin{eqnarray}
  \label{eq:f1s}
F_1^{(2)S}(q^2)&=&-\int \frac{d^3p}{(2\pi)^3} \bar\Psi_F\left({\bf
    p}+\frac{2\epsilon_d}{E_F+M_F}{\bf
    \Delta}\right)\sqrt{\frac{\epsilon_Q(p)+m_Q}{2\epsilon_Q(p)}}\sqrt{\frac{\epsilon_q(\Delta)+m_q}{2\epsilon_q(\Delta)}}\cr
&&\times\Biggl\{\frac{1}{2\epsilon_Q(\Delta)(\epsilon_Q(\Delta)+m_Q)}\Biggl[\epsilon_Q(\Delta)-m_Q+(E_F-M_F)\left(1-\frac{\epsilon_d}{E_F+M_F}\right)\cr
&&-\frac{\bf p
  \Delta}{E_F+M_F}\Biggr]\left[M_I-\epsilon_Q(p)-\epsilon_d(p)\right]
+\frac{1}{2\epsilon_q(\Delta)(\epsilon_q(\Delta)+m_q)}\Biggl[\epsilon_q(\Delta)-m_q\cr
&&+(E_F-M_F)\left(1-\frac{\epsilon_d}{E_F+M_F}\right)+\frac{\bf p
  \Delta}{E_F+M_F}\Biggr]\Biggl[M_F-\epsilon_q\left({\bf
    p}+\frac{2\epsilon_d}{E_F+M_F}{\bf
    \Delta}\right) \cr
&&-\epsilon_d\left({\bf  p}+\frac{2\epsilon_d}{E_F+M_F}{\bf\Delta}\right)\Biggr]\Biggr\}\Psi_I({\bf p});
\end{eqnarray}
\begin{eqnarray}
  \label{eq:f1v}
F_1^{(2)V}(q^2)&=&-\int \frac{d^3p}{(2\pi)^3} \bar\Psi_F\left({\bf
    p}+\frac{2\epsilon_d}{E_F+M_F}{\bf
    \Delta}\right)\sqrt{\frac{\epsilon_Q(p)+m_Q}{2\epsilon_Q(p)}}\sqrt{\frac{\epsilon_q(\Delta)+m_q}{2\epsilon_q(\Delta)}}\cr
&&\times\Biggl\{\frac{1}{2\epsilon_Q(\Delta)(\epsilon_Q(\Delta)+m_Q)}\Biggl(\left[\epsilon_Q(\Delta)-m_Q+(E_F-M_F)\left(1-\frac{\epsilon_d}{E_F+M_F}\right)\right]\cr
&&\times\left[1-\frac{\epsilon_d}{m_Q}\frac{E_F-M_F}{E_F+M_F}-\frac{\bf p
  \Delta}{E_d(E_F+M_F)}\right]-\frac{\bf p
  \Delta}{E_F+M_F}\Biggr)\left[M_I-\epsilon_Q(p)-\epsilon_d(p)\right]\cr
&&
+\frac{1}{2\epsilon_q(\Delta)(\epsilon_q(\Delta)+m_q)}\Biggl(\left[\epsilon_q(\Delta)-m_q+(E_F-M_F)\left(1-\frac{\epsilon_d}{E_F+M_F}\right)\right]\cr
&&\times\left[1-\frac{\epsilon_d}{m_q}\frac{E_F-M_F}{E_F+M_F}-\frac{\bf p
  \Delta}{E_d(E_F+M_F)}\right]-\frac{\bf p
  \Delta}{E_F+M_F}\Biggr)\cr
&&\times\Biggl[M_F-\epsilon_q\left({\bf
    p}+\frac{2\epsilon_d}{E_F+M_F}{\bf
    \Delta}\right) -\epsilon_d\left({\bf  p}+\frac{2\epsilon_d}{E_F+M_F}{\bf\Delta}\right)\Biggr]\Biggr\}\Psi_I({\bf p});
\end{eqnarray}

\begin{equation}
F_2(q^2)=F_2^{(1)}(q^2)+\varepsilon
F_2^{(2)S}(q^2)+(1-\varepsilon)F_2^{(2)V}(q^2),
\end{equation}
\begin{eqnarray}
\label{eq:f2}
F_2^{(1)}(q^2)&=&-\int \frac{d^3p}{(2\pi)^3} \bar\Psi_F\left({\bf
    p}+\frac{2\epsilon_d}{E_F+M_F}{\bf
    \Delta}\right)\sqrt{\frac{\epsilon_Q(p)+m_Q}{2\epsilon_Q(p)}}\sqrt{\frac{\epsilon_q(p+\Delta)+m_q}{2\epsilon_q(p+\Delta)}}\frac{2M_F}{E_F+M_F}\cr
&&\times\Biggl\{\frac{\epsilon_d}{\epsilon_q(p+\Delta)+m_q}\left[1+\frac{\epsilon_d}{\epsilon_Q(p)+m_Q}\frac{E_F-M_F}{E_F+M_F}\right]-\frac23\frac{{\bf
    p}^2}{(\epsilon_q(p+\Delta)+m_q)(\epsilon_Q(p)+m_Q)}\cr
&&-\frac{\bf p
  \Delta}{M_F(\epsilon_q(p+\Delta)+m_q)}\left[1-\frac12\frac{\epsilon_d}{\epsilon_Q(p)+m_Q}\frac{E_F-M_F}{E_F+M_F}\right]+\frac{\bf
  p \Delta}{{\bf \Delta}^2}\frac{E_F}{M_F}\frac{E_F+M_F}{\epsilon_Q(p)+m_Q}\cr
&& \times\left[1+\frac{\epsilon_d}{\epsilon_q(p+\Delta)+m_q}\frac{E_F-M_F}{E_F+M_F}\right]
\Biggr\}\Psi_I({\bf p});
\end{eqnarray}

\begin{eqnarray}
  \label{eq:f2s}
F_2^{(2)S}(q^2)&=&-\int \frac{d^3p}{(2\pi)^3} \bar\Psi_F\left({\bf
    p}+\frac{2\epsilon_d}{E_F+M_F}{\bf
    \Delta}\right)\sqrt{\frac{\epsilon_Q(p)+m_Q}{2\epsilon_Q(p)}}\sqrt{\frac{\epsilon_q(\Delta)+m_q}{2\epsilon_q(\Delta)}}\
\frac{\bf p \Delta}{{\bf \Delta}^2}\cr
&&\times\frac{E_F}{2\epsilon_Q(\Delta)(\epsilon_Q(\Delta)+m_Q)}\Biggl[M_I-\epsilon_Q(p)-\epsilon_d(p)+\cr
&&M_F-\epsilon_q\left({\bf
    p}+\frac{2\epsilon_d}{E_F+M_F}{\bf
    \Delta}\right) -\epsilon_d\left({\bf  p}+\frac{2\epsilon_d}{E_F+M_F}{\bf\Delta}\right)\Biggl]\Psi_I({\bf p});
\end{eqnarray}

\begin{eqnarray}
  \label{eq:f2v}
F_2^{(2)V}(q^2)&=&-\int \frac{d^3p}{(2\pi)^3} \bar\Psi_F\left({\bf
    p}+\frac{2\epsilon_d}{E_F+M_F}{\bf
    \Delta}\right)\sqrt{\frac{\epsilon_Q(p)+m_Q}{2\epsilon_Q(p)}}\sqrt{\frac{\epsilon_q(\Delta)+m_q}{2\epsilon_q(\Delta)}}\cr
&&\times
\Biggl\{\frac{\bf p \Delta}{{\bf \Delta}^2}\frac{E_F}{2\epsilon_Q(\Delta)(\epsilon_Q(\Delta)+m_Q)}\Biggl[M_I-\epsilon_Q(p)-\epsilon_d(p)\cr
&&+M_F-\epsilon_q\left({\bf
    p}+\frac{2\epsilon_d}{E_F+M_F}{\bf
    \Delta}\right) -\epsilon_d\left({\bf
    p}+\frac{2\epsilon_d}{E_F+M_F}{\bf\Delta}\right)\Biggl]+\frac{E_F-M_F}{E_F+M_F}\cr
&&\times\Biggl(\frac{\epsilon_d}{m_Q}\frac{\epsilon_Q(\Delta)-m_Q+(E_F-M_F)\left(1-\frac{\epsilon_d}{E_F+M_F}\right)}{\epsilon_Q(\Delta)(\epsilon_Q(\Delta)+m_Q)}\left[M_I-\epsilon_Q(p)-\epsilon_d(p)\right]\cr
&&+\frac{\epsilon_d}{m_q}\frac{\epsilon_q(\Delta)-m_q+(E_F-M_F)\left(1-\frac{\epsilon_d}{E_F+M_F}\right)}{\epsilon_q(\Delta)(\epsilon_q(\Delta)+m_q)}\Biggl[M_F-\epsilon_q\left({\bf
    p}+\frac{2\epsilon_d}{E_F+M_F}{\bf
    \Delta}\right)\cr
&& -\epsilon_d\left({\bf  p}+\frac{2\epsilon_d}{E_F+M_F}{\bf\Delta}\right)\Biggl]\Biggr\}
\Psi_I({\bf p});
\end{eqnarray}

\begin{equation}
F_3(q^2)=F_3^{(1)}(q^2)+\varepsilon
F_3^{(2)S}(q^2)+(1-\varepsilon)F_3^{(2)V}(q^2),
\end{equation}

\begin{eqnarray}
\label{eq:f3}
F_3^{(1)}(q^2)&=&-\int \frac{d^3p}{(2\pi)^3} \bar\Psi_F\left({\bf
    p}+\frac{2\epsilon_d}{E_F+M_F}{\bf
    \Delta}\right)\sqrt{\frac{\epsilon_Q(p)+m_Q}{2\epsilon_Q(p)}}\sqrt{\frac{\epsilon_q(p+\Delta)+m_q}{2\epsilon_q(p+\Delta)}}\frac{2M_F}{E_F+M_F}\cr
&&\times\Biggl\{\frac{\epsilon_d}{\epsilon_Q(p)+m_Q}\left[1+\frac{\epsilon_d}{\epsilon_q(p+\Delta)+m_q}\frac{E_F-M_F}{E_F+M_F}+\frac{2{\bf p
  \Delta}}{(E_F+M_F)(\epsilon_q(p+\Delta)+m_q)}\right]\cr
&&-\frac23\frac{{\bf
    p}^2}{(\epsilon_q(p+\Delta)+m_q)(\epsilon_Q(p)+m_Q)}\cr
&& -\frac{\bf
  p \Delta}{{\bf \Delta}^2}\frac{E_F+M_F}{\epsilon_Q(p)+m_Q}\left[1+\frac{\epsilon_d}{\epsilon_q(p+\Delta)+m_q}\frac{E_F-M_F}{E_F+M_F}\right]
\Biggr\}\Psi_I({\bf p});
\end{eqnarray}
\begin{eqnarray}
  \label{eq:f3s}
F_3^{(2)S}(q^2)=F_3^{(2)V}(q^2)&=&\int \frac{d^3p}{(2\pi)^3} \bar\Psi_F\left({\bf
    p}+\frac{2\epsilon_d}{E_F+M_F}{\bf
    \Delta}\right)\sqrt{\frac{\epsilon_Q(p)+m_Q}{2\epsilon_Q(p)}}\sqrt{\frac{\epsilon_q(\Delta)+m_q}{2\epsilon_q(\Delta)}}\cr
&&\times
\frac{\bf p \Delta}{{\bf \Delta}^2}\frac{M_F}{2\epsilon_Q(\Delta)(\epsilon_Q(\Delta)+m_Q)}\Biggl[M_I-\epsilon_Q(p)-\epsilon_d(p)+M_F\cr
&&-\epsilon_q\left({\bf
    p}+\frac{2\epsilon_d}{E_F+M_F}{\bf
    \Delta}\right) -\epsilon_d\left({\bf  p}+\frac{2\epsilon_d}{E_F+M_F}{\bf\Delta}\right)\Biggl]\Psi_I({\bf p}).\qquad
\end{eqnarray}

b) Axial vector form factors
\begin{equation}
G_1(q^2)=G_1^{(1)}(q^2)+\varepsilon
G_1^{(2)S}(q^2)+(1-\varepsilon)G_1^{(2)V}(q^2),
\end{equation}

\begin{eqnarray}
\label{eq:g1}
G_1^{(1)}(q^2)&=&\int \frac{d^3p}{(2\pi)^3} \bar\Psi_F\left({\bf
    p}+\frac{2\epsilon_d}{E_F+M_F}{\bf
    \Delta}\right)\sqrt{\frac{\epsilon_Q(p)+m_Q}{2\epsilon_Q(p)}}\sqrt{\frac{\epsilon_q(p+\Delta)+m_q}{2\epsilon_q(p+\Delta)}}\cr
&&\times
\Biggl\{1+\frac{E_F-M_F}{E_F+M_F}\Biggl(\frac{\epsilon_d}{\epsilon_q(p+\Delta)+m_q}\left[1+\frac{\epsilon_d}{\epsilon_Q(p)+m_Q}\right]+\frac{\epsilon_d}{\epsilon_Q(p)+m_Q}\Biggr)\cr
&&-\frac13\frac{{\bf
    p}^2}{(\epsilon_q(p+\Delta)+m_q)(\epsilon_Q(p)+m_Q)}+ \frac{\bf p
  \Delta}{E_F+M_F}\Biggl[\frac1{\epsilon_q(p+\Delta)+m_q}\cr
&&-\frac1{\epsilon_Q(p)+m_Q}-\frac{2\epsilon_d}{(\epsilon_q(p+\Delta)+m_q)(\epsilon_Q(p)+m_Q)}\Biggr]\Biggr\}\Psi_I({\bf
  p}); 
\end{eqnarray}

\begin{eqnarray}
  \label{eq:g1s}
G_1^{(2)S}(q^2)&=&-\int \frac{d^3p}{(2\pi)^3} \bar\Psi_F\left({\bf
    p}+\frac{2\epsilon_d}{E_F+M_F}{\bf
    \Delta}\right)\sqrt{\frac{\epsilon_Q(p)+m_Q}{2\epsilon_Q(p)}}\sqrt{\frac{\epsilon_q(\Delta)+m_q}{2\epsilon_q(\Delta)}}\cr
&&\times\Biggl\{\frac{1}{2\epsilon_Q(\Delta)(\epsilon_Q(\Delta)+m_Q)}\Biggl[\epsilon_Q(\Delta)-m_Q+(E_F-M_F)\left(1-\frac{\epsilon_d}{E_F+M_F}\right)\cr
&&+\frac{\bf p
  \Delta}{E_F+M_F}\Biggr]\left[M_I-\epsilon_Q(p)-\epsilon_d(p)\right]
+\frac{1}{2\epsilon_q(\Delta)(\epsilon_q(\Delta)+m_q)}\Biggl[\epsilon_q(\Delta)-m_q\cr
&&+(E_F-M_F)\left(1-\frac{\epsilon_d}{E_F+M_F}\right)+\frac{\bf p
  \Delta}{E_F+M_F}\Biggr]\Biggl[M_F-\epsilon_q\left({\bf
    p}+\frac{2\epsilon_d}{E_F+M_F}{\bf
    \Delta}\right) \cr
&&-\epsilon_d\left({\bf  p}+\frac{2\epsilon_d}{E_F+M_F}{\bf\Delta}\right)\Biggr]\Biggr\}\Psi_I({\bf p});
\end{eqnarray}

\begin{eqnarray}
  \label{eq:g1v}
G_1^{(2)V}(q^2)&=&-\int \frac{d^3p}{(2\pi)^3} \bar\Psi_F\left({\bf
    p}+\frac{2\epsilon_d}{E_F+M_F}{\bf
    \Delta}\right)\sqrt{\frac{\epsilon_Q(p)+m_Q}{2\epsilon_Q(p)}}\sqrt{\frac{\epsilon_q(\Delta)+m_q}{2\epsilon_q(\Delta)}}\cr
&&\times\Biggl\{\frac{1}{2\epsilon_Q(\Delta)(\epsilon_Q(\Delta)+m_Q)}\Biggl(\left[\epsilon_Q(\Delta)-m_Q+(E_F-M_F)\left(1-\frac{\epsilon_d}{E_F+M_F}\right)\right]\cr
&&\times\left[1+\frac{\epsilon_d}{m_Q}\frac{E_F-M_F}{E_F+M_F}-\frac{\bf p
  \Delta}{E_d(E_F+M_F)}\right]+\frac{\bf p
  \Delta}{E_F+M_F}\Biggr)\left[M_I-\epsilon_Q(p)-\epsilon_d(p)\right]\cr
&&
+\frac{1}{2\epsilon_q(\Delta)(\epsilon_q(\Delta)+m_q)}\Biggl(\left[\epsilon_q(\Delta)-m_q+(E_F-M_F)\left(1-\frac{\epsilon_d}{E_F+M_F}\right)\right]\cr
&&\times\left[1-\frac{\epsilon_d}{m_q}\frac{E_F-M_F}{E_F+M_F}-\frac{\bf p
  \Delta}{E_d(E_F+M_F)}\right]-\frac{\bf p
  \Delta}{E_F+M_F}\Biggr)\cr
&&\times\Biggl[M_F-\epsilon_q\left({\bf
    p}+\frac{2\epsilon_d}{E_F+M_F}{\bf
    \Delta}\right) -\epsilon_d\left({\bf  p}+\frac{2\epsilon_d}{E_F+M_F}{\bf\Delta}\right)\Biggr]\Biggr\}\Psi_I({\bf p});
\end{eqnarray}

\begin{equation}
G_2(q^2)=G_2^{(1)}(q^2)+\varepsilon
G_2^{(2)S}(q^2)+(1-\varepsilon)G_2^{(2)V}(q^2),
\end{equation}

\begin{eqnarray}
\label{eq:g2}
G_2^{(1)}(q^2)&=&-\int \frac{d^3p}{(2\pi)^3} \bar\Psi_F\left({\bf
    p}+\frac{2\epsilon_d}{E_F+M_F}{\bf
    \Delta}\right)\sqrt{\frac{\epsilon_Q(p)+m_Q}{2\epsilon_Q(p)}}\sqrt{\frac{\epsilon_q(p+\Delta)+m_q}{2\epsilon_q(p+\Delta)}}\frac{2M_F}{E_F+M_F}\cr
&&\times\Biggl\{\frac{\epsilon_d}{\epsilon_q(p+\Delta)+m_q}\left[1+\frac{\epsilon_d}{\epsilon_Q(p)+m_Q}\right]-\frac{\bf p
  \Delta}{M_F(\epsilon_q(p+\Delta)+m_q)}\cr
&& \times\Biggl[1-\frac{\epsilon_d}{\epsilon_Q(p)+m_Q}\frac{M_F}{E_F+M_F}\Biggr]-\frac{\bf
  p \Delta}{{\bf \Delta}^2}\frac{E_F}{M_F}\frac{E_F+M_F}{\epsilon_Q(p)+m_Q}\cr
&& \times\left[1+\frac{\epsilon_d}{\epsilon_q(p+\Delta)+m_q}\right]
\Biggr\}\Psi_I({\bf p});
\end{eqnarray}

\begin{eqnarray}
  \label{eq:g2s}
G_2^{(2)S}(q^2)&=&-\int \frac{d^3p}{(2\pi)^3} \bar\Psi_F\left({\bf
    p}+\frac{2\epsilon_d}{E_F+M_F}{\bf
    \Delta}\right)\sqrt{\frac{\epsilon_Q(p)+m_Q}{2\epsilon_Q(p)}}\sqrt{\frac{\epsilon_q(\Delta)+m_q}{2\epsilon_q(\Delta)}}\cr
&&\times\Biggl\{\frac{\bf p
  \Delta}{E_F+M_F}\Biggl(\frac{1}{\epsilon_Q(\Delta)(\epsilon_Q(\Delta)+m_Q)}\left[M_I-\epsilon_Q(p)-\epsilon_d(p)\right]\cr
&&+\frac{1}{\epsilon_q(\Delta)(\epsilon_q(\Delta)+m_q)}\Biggl[M_F-\epsilon_q\left({\bf
    p}+\frac{2\epsilon_d}{E_F+M_F}{\bf
    \Delta}\right) -\epsilon_d\left({\bf
    p}+\frac{2\epsilon_d}{E_F+M_F}{\bf\Delta}\right)\Biggl]\Biggr)\cr
&&-
\frac{\bf p \Delta}{{\bf \Delta}^2}\frac{E_F}{2\epsilon_Q(\Delta)(\epsilon_Q(\Delta)+m_Q)}\Biggl[M_I-\epsilon_Q(p)-\epsilon_d(p)\cr
&&+M_F-\epsilon_q\left({\bf
    p}+\frac{2\epsilon_d}{E_F+M_F}{\bf
    \Delta}\right) -\epsilon_d\left({\bf  p}+\frac{2\epsilon_d}{E_F+M_F}{\bf\Delta}\right)\Biggl]\Psi_I({\bf p});
\end{eqnarray}

\begin{eqnarray}
  \label{eq:g2v}
G_2^{(2)V}(q^2)&=&-\int \frac{d^3p}{(2\pi)^3} \bar\Psi_F\left({\bf
    p}+\frac{2\epsilon_d}{E_F+M_F}{\bf
    \Delta}\right)\sqrt{\frac{\epsilon_Q(p)+m_Q}{2\epsilon_Q(p)}}\sqrt{\frac{\epsilon_q(\Delta)+m_q}{2\epsilon_q(\Delta)}}\cr
&&\times\Biggl\{\frac{1}{\epsilon_Q(\Delta)(\epsilon_Q(\Delta)+m_Q)}\Biggl(\Biggl[\epsilon_Q(\Delta)-m_Q+(E_F-M_F)\left(1-\frac{\epsilon_d}{E_F+M_F}\right)\Biggr]\cr
&&\times\left[\frac{\bf p \Delta}{E_d(E_F+M_F)}-\frac{\epsilon_d}{m_Q}\frac{E_F-M_F}{E_F+M_F}\right]-\frac{\bf p
  \Delta}{E_F+M_F}\Biggr)\left[M_I-\epsilon_Q(p)-\epsilon_d(p)\right]\cr
&&
+\frac{1}{\epsilon_q(\Delta)(\epsilon_q(\Delta)+m_q)}\Biggl(\Biggl[\epsilon_q(\Delta)-m_q-(E_F-M_F)\left(1-\frac{\epsilon_d}{E_F+M_F}\right)\Biggr]\cr
&&\times\left[-\frac{\bf p \Delta}{E_d(E_F+M_F)}+\frac{\epsilon_d}{m_q}\frac{E_F-M_F}{E_F+M_F}\right]+\frac{\bf p
  \Delta}{E_F+M_F}\Biggr)\Biggl[M_F-\epsilon_q\left({\bf
    p}+\frac{2\epsilon_d}{E_F+M_F}{\bf
    \Delta}\right) \cr
&&-\epsilon_d\left({\bf  p}+\frac{2\epsilon_d}{E_F+M_F}{\bf\Delta}\right)\Biggr]-\frac{\bf p \Delta}{{\bf \Delta}^2}\frac{E_F}{2\epsilon_Q(\Delta)(\epsilon_Q(\Delta)+m_Q)}\Biggl[M_I-\epsilon_Q(p)-\epsilon_d(p)\cr
&&+M_F-\epsilon_q\left({\bf
    p}+\frac{2\epsilon_d}{E_F+M_F}{\bf
    \Delta}\right) -\epsilon_d\left({\bf
    p}+\frac{2\epsilon_d}{E_F+M_F}{\bf\Delta}\right)\Biggl]\Biggr\}\Psi_I({\bf p});
\end{eqnarray}

\begin{equation}
G_3(q^2)=G_3^{(1)}(q^2)+\varepsilon
G_3^{(2)S}(q^2)+(1-\varepsilon)G_3^{(2)V}(q^2),
\end{equation}

\begin{eqnarray}
\label{eq:g3}
G_3^{(1)}(q^2)&=&\int \frac{d^3p}{(2\pi)^3} \bar\Psi_F\left({\bf
    p}+\frac{2\epsilon_d}{E_F+M_F}{\bf
    \Delta}\right)\sqrt{\frac{\epsilon_Q(p)+m_Q}{2\epsilon_Q(p)}}\sqrt{\frac{\epsilon_q(p+\Delta)+m_q}{2\epsilon_q(p+\Delta)}}\frac{2M_F}{E_F+M_F}\cr
&&\times\Biggl\{\frac{\epsilon_d}{\epsilon_Q(p)+m_Q}\left[1+\frac{\epsilon_d}{\epsilon_q(p+\Delta)+m_q}-\frac{{\bf p
  \Delta}}{(E_F+M_F)(\epsilon_q(p+\Delta)+m_q)}\right]\cr
&& -\frac{\bf
  p \Delta}{{\bf \Delta}^2}\frac{E_F+M_F}{\epsilon_Q(p)+m_Q}\left[1+\frac{\epsilon_d}{\epsilon_q(p+\Delta)+m_q}\right]
\Biggr\}\Psi_I({\bf p});
\end{eqnarray}
\begin{eqnarray}
  \label{eq:g3s}
G_3^{(2)S}(q^2)=G_3^{(2)V}(q^2)&=&-\int \frac{d^3p}{(2\pi)^3} \bar\Psi_F\left({\bf
    p}+\frac{2\epsilon_d}{E_F+M_F}{\bf
    \Delta}\right)\sqrt{\frac{\epsilon_Q(p)+m_Q}{2\epsilon_Q(p)}}\sqrt{\frac{\epsilon_q(\Delta)+m_q}{2\epsilon_q(\Delta)}}\cr
&&\times
\frac{\bf p \Delta}{{\bf \Delta}^2}\frac{M_F}{2\epsilon_Q(\Delta)(\epsilon_Q(\Delta)+m_Q)}\Biggl[M_I-\epsilon_Q(p)-\epsilon_d(p)+M_F\cr
&&-\epsilon_q\left({\bf
    p}+\frac{2\epsilon_d}{E_F+M_F}{\bf
    \Delta}\right) -\epsilon_d\left({\bf  p}+\frac{2\epsilon_d}{E_F+M_F}{\bf\Delta}\right)\Biggl]\Psi_I({\bf p});
\end{eqnarray}
where 
\[ \left|{\bf \Delta}\right|=\sqrt{\frac{(M_{I}^2+M_F^2-q^2)^2}
{4M_{I}^2}-M_F^2},\]
superscripts (1) and (2) correspond to vertex functions $\Gamma^{(1)}$
and $\Gamma^{(2)}$,  $S$ and $V$ correspond to the scalar and vector confining
potentials, $\epsilon_d$ is the diquark energy,
\[ E_F=\sqrt{M_F^2+{\bf \Delta}^2},\quad \epsilon_q(
\Delta)=\sqrt{m_q^2+{\bf\Delta}^2}, \quad
 \epsilon_q(p+\lambda
\Delta)=\sqrt{m_q^2+({\bf p}+\lambda{\bf \Delta})^2} \quad (q=b,c,u,d), \]
subscripts $I$ and $F$ denote the initial $\Lambda_Q$ and final $\Lambda_q$ baryons,
and  the subscript $q$ corresponds to $c$ or $u$ quark for the final
$\Lambda_c$ or $p$, respectively.


\begin{thebibliography}{00}
\bibitem{pdg}  K.A. Olive {\it et al.} (Particle Data Group), 
 ``Review of particle physics,''  
Chin. Phys. C, {\bf 38}, 090001 (2014).
\bibitem{belle} 
  Y.~Sato {\it et al.} [Belle Collaboration],
  ``Measurement of the branching ratio of $\bar{B}^0 \rightarrow D^{*+} \tau^- \bar{\nu}_{\tau}$ relative to $\bar{B}^0 \rightarrow D^{*+} \ell^- \bar{\nu}_{\ell}$ decays with a semileptonic tagging method,''
  arXiv:1607.07923 [hep-ex].
\bibitem{kkhm} 
  X.~W.~Kang, B.~Kubis, C.~Hanhart and Ulf-G.~Meissner,
  ``$B_{l4}$ decays and the extraction of $|V_{ub}|$,''
  Phys.\ Rev.\ D {\bf 89}, 053015 (2014).
\bibitem{lhcb} 
  R.~Aaij {\it et al.} [LHCb Collaboration],
  ``Determination of the quark coupling strength $|V_{ub}|$ using baryonic decays,''
  Nature Phys.\  {\bf 11}, 743 (2015).
  \bibitem{mass} D.~Ebert, V.~O.~Galkin and R.~N.~Faustov,
  ``Mass spectrum of orbitally and radially excited heavy - light mesons in the relativistic quark model,''
  Phys.\ Rev.\ D {\bf 57}, 5663 (1998)
  [Phys.\ Rev.\ D {\bf 59}, 019902 (1999)];
   ``Properties of heavy quarkonia and $B_c$ mesons in the relativistic quark model,''
  Phys. Rev. D {\bf 67}, 014027 (2003).
  \bibitem{slbdecay} 
  D.~Ebert, R.~N.~Faustov and V.~O.~Galkin,
  ``Analysis of semileptonic $B$ decays in the relativistic quark model,''
  Phys.\ Rev.\ D {\bf 75}, 074008 (2007);
R.~N.~Faustov and V.~O.~Galkin,
  ``Exclusive weak $B$ decays involving $\tau$ lepton in the relativistic quark model,''
  Mod.\ Phys.\ Lett.\ A {\bf 27}, 1250183 (2012);
``Rare $B\to \pi l\bar l$ and $B\to\rho l \bar l$ decays in the relativistic quark model,''
  Eur.\ Phys.\ J.\ C {\bf 74}, no. 6, 2911 (2014).  
\bibitem{hbar} 
  D.~Ebert, R.~N.~Faustov and V.~O.~Galkin,
  ``Masses of heavy baryons in the relativistic quark model,''
  Phys.\ Rev.\ D {\bf 72}, 034026 (2005);
``Masses of excited heavy baryons in the relativistic quark model,''
  Phys.\ Lett.\ B {\bf 659}, 612 (2008).
  \bibitem{barregge} 
  D.~Ebert, R.~N.~Faustov and V.~O.~Galkin,
  ``Spectroscopy and Regge trajectories of heavy baryons in the relativistic quark-diquark picture,''
  Phys.\ Rev.\ D {\bf 84}, 014025 (2011); R.~N.~Faustov and V.~O.~Galkin,
  ``Strange baryon spectroscopy in the relativistic quark model,''
  Phys.\ Rev.\ D {\bf 92}, no. 5, 054005 (2015).
  \bibitem{hbardecay} 
  D.~Ebert, R.~N.~Faustov and V.~O.~Galkin,
  ``Semileptonic decays of heavy baryons in the relativistic quark model,''
  Phys.\ Rev.\ D {\bf 73}, 094002 (2006).
  \bibitem{iwb} 
  N.~Isgur and M.~B.~Wise,
  ``Heavy baryon weak form-factors,''
  Nucl.\ Phys.\ B {\bf 348}, 276 (1991); A.~F.~Falk and M.~Neubert,
  ``Second order power corrections in the heavy quark effective theory. 2. Baryon form-factors,''
  Phys.\ Rev.\ D {\bf 47}, 2982 (1993).


 
  \bibitem{ltetr} 
  D.~Ebert, R.~N.~Faustov and V.~O.~Galkin,
  ``Mass spectra and Regge trajectories of light mesons in the relativistic quark model,''
  Phys.\ Rev.\ D {\bf 79}, 114029 (2009);
   ``Masses of light tetraquarks and scalar mesons in the relativistic quark model,''
  Eur.\ Phys.\ J.\ C {\bf 60}, 273 (2009).
\bibitem{f} R.~N.~Faustov,
  ``Relativistic wave function and form-factors of the bound system,''
  Annals Phys.\  {\bf 78}, 176 (1973); 
  ``Magnetic moment of the relativistic composite system,''
  Nuovo Cim.\ A {\bf 69}, 37 (1970). 
\bibitem{giklsh} 
T.~Gutsche, M.~A.~Ivanov, J.~G.~K\"orner, V.~E.~Lyubovitskij, P.~Santorelli and N.~Habyl,
  ``Semileptonic decay $\Lambda_b \to \Lambda_c + \tau^- + \bar{\nu_\tau}$ in the covariant confined quark model,''
  Phys.\ Rev.\ D {\bf 91}, no. 7, 074001 (2015)
  Erratum: [Phys.\ Rev.\ D {\bf 91}, no. 11, 119907 (2015)].
\bibitem{gikls} 
  T.~Gutsche, M.~A.~Ivanov, J.~G.~K\"orner, V.~E.~Lyubovitskij and P.~Santorelli,
  ``Heavy-to-light semileptonic decays of $\Lambda_b$ and $\Lambda_c$ baryons in the covariant confined quark model,''
  Phys.\ Rev.\ D {\bf 90}, no. 11, 114033 (2014);
``Semileptonic decays $\Lambda_c^+ \to \Lambda \ell^+ \nu_\ell\,\,(\ell=e,\mu)$ in the covariant quark model and comparison with the new absolute branching fraction measurements of Belle and BESIII,''
  Phys.\ Rev.\ D {\bf 93}, no. 3, 034008 (2016).
\bibitem{hmdecay} 
  R.~N.~Faustov and V.~O.~Galkin,
  ``Heavy quark $1/m_Q$ expansion of meson weak decay form-factors in the relativistic quark model,''
  Z.\ Phys.\ C {\bf 66}, 119 (1995).
\bibitem{prc} 
  M.~Pervin, W.~Roberts and S.~Capstick,
  ``Semileptonic decays of heavy lambda baryons in a quark model,''
  Phys.\ Rev.\ C {\bf 72}, 035201 (2005).
\bibitem{feldmann} 
  T.~Feldmann and M.~W.~Y.~Yip,
  ``Form factors for $\Lambda_b \to \Lambda$ transitions in {SCET},''
  Phys.\ Rev.\ D {\bf 85}, 014035 (2012) Erratum: [Phys.\ Rev.\ D {\bf
    86}, 079901 (2012)]; P.~Böer, T.~Feldmann and D.~van Dyk,
  ``Angular analysis of the decay $\Lambda_b \to \Lambda (\to N \pi) \ell^+\ell^-$,''
  JHEP {\bf 1501}, 155 (2015).

  \bibitem{wkl} 
  H.~W.~Ke, X.~Q.~Li and Z.~T.~Wei,
  ``Diquarks and $\Lambda_b \to \Lambda_c$ weak decays,''
  Phys.\ Rev.\ D {\bf 77}, 014020 (2008);
Z.~T.~Wei, H.~W.~Ke and X.~Q.~Li,
  ``Evaluating decay Rates and Asymmetries of $\Lambda_b$ into Light Baryons in LFQM,''
  Phys.\ Rev.\ D {\bf 80}, 094016 (2009).
  \bibitem{kkmw} 
  A.~Khodjamirian, C.~Klein, T.~Mannel and Y.-M.~Wang,
  ``Form Factors and Strong Couplings of Heavy Baryons from QCD Light-Cone Sum Rules,''
  JHEP {\bf 1109}, 106 (2011).

\bibitem{kk} 
  P.~Bialas, J.~G.~K\"orner, M.~Kramer and K.~Zalewski,
  ``Joint angular decay distributions in exclusive weak decays of heavy mesons and baryons,''
  Z.\ Phys.\ C {\bf 57}, 115 (1993).
\bibitem{dutta} 
  R.~Dutta,
  ``$\Lambda_b \to (\Lambda_c,\,p)\,\tau\,\nu$ decays within standard model and beyond,''
  Phys.\ Rev.\ D {\bf 93}, no. 5, 054003 (2016).
\bibitem{lat} 
  W.~Detmold, C.~Lehner and S.~Meinel,
  ``$\Lambda_b \to p \ell^- \bar{\nu}_\ell$ and $\Lambda_b \to \Lambda_c \ell^- \bar{\nu}_\ell$ form factors from lattice QCD with relativistic heavy quarks,''
  Phys.\ Rev.\ D {\bf 92}, no. 3, 034503 (2015).

\bibitem{swd} 
  S.~Shivashankara, W.~Wu and A.~Datta,
  ``$\Lambda_b \to \Lambda_c \tau \bar{\nu}_{\tau}$ decay in the standard model and with new physics,''
  Phys.\ Rev.\ D {\bf 91}, no. 11, 115003 (2015)


\end{thebibliography}
\end{document}